\documentclass{article}

\usepackage{microtype}
\usepackage{graphicx}
\usepackage{subfigure}
\usepackage{booktabs} 

\usepackage{array,multirow} 
\usepackage{varwidth} 
\usepackage{longtable}

\usepackage{hyperref}

\usepackage[accepted]{icml2024}

\usepackage{amsmath}
\usepackage{amssymb}
\usepackage{mathtools}
\usepackage{amsthm}
\usepackage{physics}

\usepackage[capitalize,noabbrev]{cleveref}

\theoremstyle{plain}

\theoremstyle{definition}

\theoremstyle{remark}

\usepackage[textsize=tiny]{todonotes}

\usepackage{color}
\def\shownotes{1}
\ifnum\shownotes=1
\newcommand{\authnote}[2]{[#1: #2]}
\else
\newcommand{\authnote}[2]{}
\fi

\icmltitlerunning{Meta-Designing Quantum Experiments with Language Models}

\begin{document}
\twocolumn[
\icmltitle{Meta-Designing Quantum Experiments with Language Models}



\icmlsetsymbol{equal}{*}

\begin{icmlauthorlist}
\icmlauthor{S\"{o}ren Arlt}{mpl}
\icmlauthor{Haonan Duan}{vector}
\icmlauthor{Felix Li}{berkeley}
\icmlauthor{Sang Michael Xie}{stanford}
\icmlauthor{Yuhuai Wu}{xai}
\icmlauthor{Mario Krenn}{mpl,tuebingen}
\end{icmlauthorlist}

\icmlaffiliation{mpl}{Max Planck Institute for the Science of Light}
\icmlaffiliation{tuebingen}{University of Tuebingen}
\icmlaffiliation{stanford}{Stanford University}
\icmlaffiliation{xai}{xAI}
\icmlaffiliation{berkeley}{UC Berkeley}
\icmlaffiliation{vector}{University of Toronto, Vector Institute}

\icmlcorrespondingauthor{S\"{o}ren Arlt}{soeren.arlt@mpl.mpg.de}
\icmlcorrespondingauthor{Haonan Duan}{haonand@cs.toronto.edu}
\icmlcorrespondingauthor{Yuhuai Wu}{yuhuai@x.ai}
\icmlcorrespondingauthor{Mario Krenn}{mario.krenn@mpl.mpg.de}


\vskip 0.3in
]



\printAffiliationsAndNotice{}  

\begin{abstract}
Artificial Intelligence (AI) can solve complex scientific problems beyond human capabilities, but the resulting solutions offer little insight into the underlying physical principles. One prominent example is quantum physics, where computers can discover experiments for the generation of specific quantum states, but it is unclear how finding general design concepts can be automated. Here, we address this challenge by training a transformer-based language model to create human-readable Python code, which solves an entire class of problems in a single pass. This strategy, which we call meta-design, enables scientists to gain a deeper understanding and extrapolate to larger experiments without additional optimization. To demonstrate the effectiveness of our approach, we uncover previously unknown experimental generalizations of important quantum states, e.g. from condensed matter physics. The underlying methodology of meta-design can naturally be extended to fields such as materials science or engineering.
\end{abstract}

\subsection*{Introduction}
Quantum physics is a notoriously unintuitive field of study. Despite this, it has developed to a point where some of its most difficult-to-conceptualize effects - such as entanglement - could become the basis of a new generation of technological development. These applications include quantum imaging \cite{lemos2014quantum,kviatkovsky2020microscopy,Moreau2019,Aslam2023}, quantum metrology \cite{dicke_metrology,Polino2020,DeMille2024}, quantum communication \cite{Flamini2018,Couteau2023-ql}, quantum simulation \cite{AspuruGuzik2012,Cornish2024-mg}, and quantum computation \cite{Madsen2022-ht}. Due to difficulties in designing experimental setups by hand, researchers have started to utilize algorithmic optimization and AI techniques to discover experimental setups in quantum physics \cite{Krenn2020}.
\\
AI techniques have been previously applied to the search for experimental setups in quantum physics \cite{krenn2016automated,Knott2016,Nichols2019,PRXQuantum.1.010301,prabhu2020accelerating,theseus,pytheus,Goel2024,https://doi.org/10.48550/arxiv.2404.14887}, nanophotonic structures \cite{Molesky2018,sapra2020chip,Ma2020,gedeon2023free}, and quantum circuits \cite{ostaszewski2021reinforcement, https://doi.org/10.48550/arxiv.2311.18588,Kottmann2023, https://doi.org/10.48550/arxiv.2402.17761,maclellan2024end}. In all of these works the algorithm produces only a single solution, which surpasses designs by human experts. For example, for a given target quantum state, a machine could design the experimental setup which creates the state, but interpretation and generalization of the results is left to the researcher and is often an exceptionally hard challenge, if possible at all.\\

In this work, we introduce the process of \textit{meta-design}\footnote{Our code is available on github: \url{https://github.com/artificial-scientist-lab/metadesign}}. Instead of designing one solution for a single target (i.e. one experimental setup for the creation of one quantum state), we train a sequence-to-sequence transformer to generate a meta-solution in the form of programming code. A meta-solution solves an infinitely large class of targets (a class of quantum states) by generating different experimental setups for different system sizes.\\
From a software-engineering perspective, our approach can be described as the automated discovery of \textit{meta-programs}, programs that generate or transform other programs \cite{krzysztof2000generative}. In our case, the discovered \textit{meta-solution} generates different programs that construct experimental setups for different system size parameters.\\
Transformer architectures have demonstrated remarkable success in solving a wide range of mathematics and physics reasoning tasks. \cite{lample2019deep} and \cite{charton22} show that a transformer-based sequence-to-sequence model can tackle symbolic math problems such as symbolic integration, differential equations and symbolic regression. AlphaGeometry \cite{alphageometry} has achieved remarkable performance in solving geometry problems at an olympiad level. \cite{alfarano2023discovering} finds that by training transformers on synthetic data, they can accurately predict the Lyapunov functions of polynomial and non-polynomial dynamical systems. In the field of theoretical high-energy physics, \cite{cai2024transforming} applies transformers to compute scattering amplitudes. Furthermore, \cite{alnuqaydansymbolic} demonstrates that a transformer model, when trained on symbolic sequence pairs, can correctly predict the squared amplitudes of Standard Model processes. Language models have also been used in quantum simulation \cite{melko2024language}, albeit not as a generative model for symbolic language as in our or the other works mentioned here.\\
In recent years, automated code synthesis with large language models have been very promising and powerful, and fit more naturally to our work. Building on systematic baselines such as Program Synthesis with Large Language Models \cite{austin2021program}, the field has advanced through large-scale sampling and clustering in AlphaCode \cite{li2022competition}, evolutionary search loops in FunSearch \cite{romera2024mathematical} and AlphaEvolve \cite{novikov2025alphaevolve}, and reinforcement-learning refinement in StepCoder \cite{dou2024stepcoder}.
Our technique differs from other automated code synthesis approaches in the purpose of the generated codes: Each code describes a generalization of a quantum optics experiment. For that, it has to learn the underlying physical design principles from data and apply them to output a meta-solution. The resulting Python codes can then be read by humans, who can directly learn these general design principles from the readable code. The readability of the code representation helps to uncover the underlying patterns in the class of solutions. Therefore, our technique is a step towards AI methods that can help to gain new understanding in physics \cite{de2017understanding, Krenn2022, barman2024towards}. In that sense, the meta-design idea shares technical details with the field of program synthesis, however, when applied for automated scientific discovery, it offers fundamentally new insights that are very challenging to obtain with other known techniques.

\begin{figure*}
    \centering
    \includegraphics[width = 0.98\textwidth]{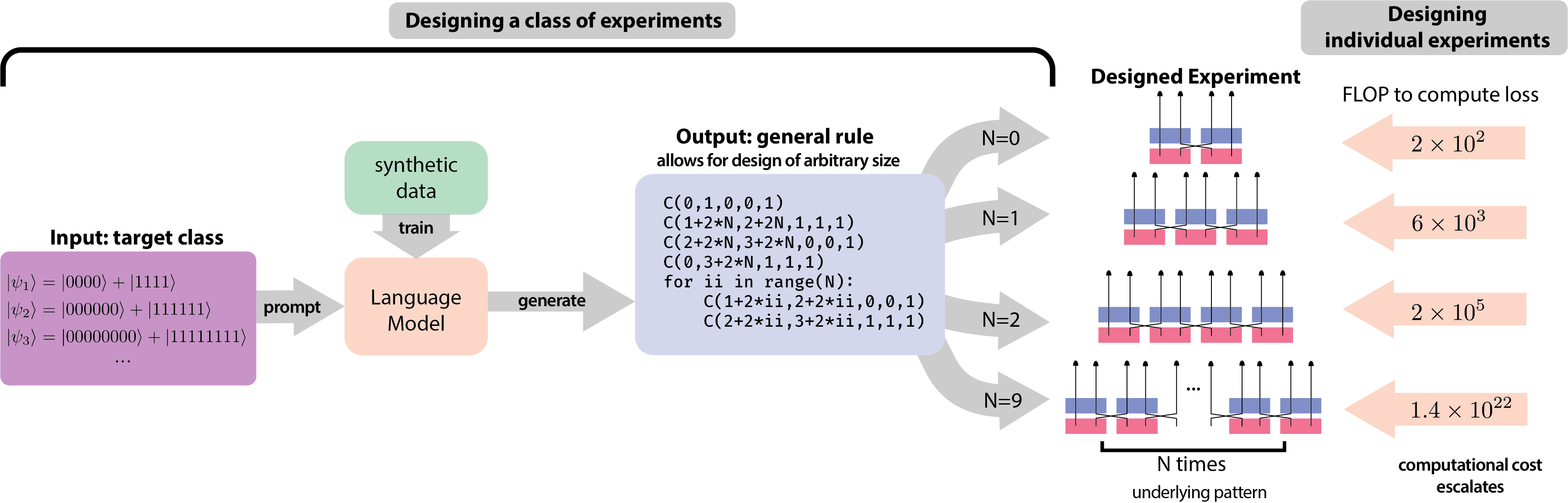}
    \caption{\textbf{Meta-designing a class of experiments via code generation avoids exploding computational costs for the design of larger experiments.} Left side:  Our process takes the first three states from a class of target quantum states and - when successful - produces a Python code which generates the correct experimental setup for arbitrary sizes. For this, a sequence-to-sequence transformer is trained purely on randomly generated pairs of sequences. Right side: Designing an experimental setup which produces a target quantum state is very fast for small particle numbers, but the computational cost explodes as the target state grows.}
    \label{fig:fig1}
\end{figure*}
\subsection*{Background Quantum Optics}
We choose the design of quantum optics experiments as proof of concept and point to the great potential in applying the approach in other fields. Quantum optics is concerned with photons, the fundamental particles of light. A photon can have different polarization modes, e.g. horizontal (mode 0) or vertical (mode 1). A basic property of quantum particles is that they can be in a superposition of multiple modes, i.e. they can be considered to be two things at the same time. The state $\ket{\psi}$ of one photon in equal superposition can be expressed in Dirac notation as
\begin{align}
    \ket{\psi} = \ket{0}+\ket{1}.
\end{align}
\begin{figure}
    \centering
    \includegraphics[width = 0.45\textwidth]{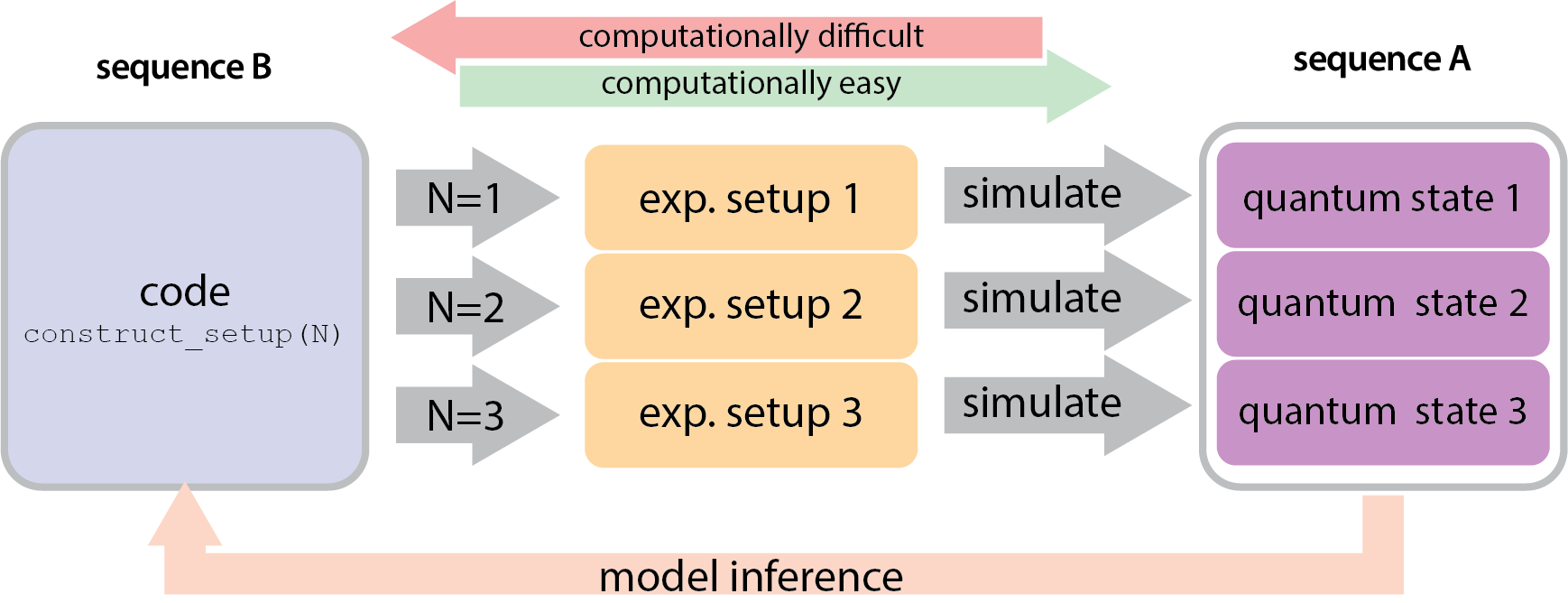}
    \caption{\textbf{Exploiting asymmetric cost for data generation.} A random Python program (sequence B) is generated. Executing it for the values $N=0,1,2$ produces three different experimental setup. Each setup produces a state. The three states are concatenated to make sequence A, which is the input for the model.}
    \label{fig:datagen_main}
\end{figure}
We omit the normalization factor for all quantum states shown in this work for readability. It can be assumed that all states are normalized. Another important concept in quantum physics is \textit{entanglement}, where multiple photons are in a state where they cannot be described independently, such as the superposition of three particles being in the superposition of either all particles being in mode 0 or all particles being in mode 1,
\begin{align}
    \ket{\psi} = \ket{000}+\ket{111}.
\end{align}
This state is called the GHZ state \cite{Greenberger1990,Pan2000}.\\
In quantum optics, highly entangled states can be created by combining probabilistic photon pair sources. The number of possible experimental setups increases combinatorically with the number of photons for a given target state. This makes it very difficult to design experiments by hand and thus computational techniques have been successfully applied to the problem \cite{pytheus}. For sufficiently large systems, these tasks become too difficult even for current methods as they become too computationally expensive (see right side of Fig. \ref{fig:fig1}).
\subsection*{Methods}
We introduce meta-design, the idea of generating a single meta-solution which can solve a whole series of problems (in our case, for design problems of quantum states of increasing particle number). We define meta-solutions as Python programs that can generate blueprints of multiple experimental setups (see Fig. \ref{fig:fig1}), e.g. a function \texttt{construct\_setup(N)}, which constructs valid setups for a growing system size $N=0,1,2...$. In the quantum optics example the code makes calls to a function defined in the software package PyTheus to generate and simulate the setups \cite{pytheus}. There is no explicit limitation on the syntax or language of the script used. We train a sequence-to-sequence transformer on synthetic data to translate from a class of quantum states to Python code and sample the model to discover meta-solutions for a selection of twenty classes of quantum states.

\paragraph{Meta-design for Quantum Experiments}
A famous class of quantum states are the GHZ states, which are shown in the left box of Fig. \ref{fig:fig1}. They are superposition of particles being either in mode 0 or mode 1 with an increasing number of photons (4, 6, 8, ...). Because the experimental setups are based on pair-sources, only an even number of photons can be created.
We now aim to find a program \texttt{construct\_setup(N)} which generates the correct experimental setup for creating GHZ states for a given particle number $2N+4$. This is possible because the GHZ states follow a specific pattern. The solution to the problem is shown in Fig. \ref{fig:fig1}.  After constructing the setup, we can compute the expected quantum state that emerges at the detectors.\\
The code shown in Fig. \ref{fig:fig1} will generate the correct experimental setups for arbitrarily high particle numbers. This code can express the correct experiments for an entire class of quantum states. The goal of this work is to show that it is possible to use language models for the discovery of programs, which solve classes of quantum states such as the GHZ state.
\paragraph{Synthetic data generation}
On an abstract level, we can describe the subject of our work as dealing with two sequences, \texttt{A} (a list of three quantum states) and \texttt{B} (python program). Direction \texttt{B→A} (computing the resulting quantum states from experimental setups) follows clear instructions and can be considered \textit{easy}. Direction \texttt{A→B} is highly non-trivial. Designing experiments can be very difficult because it usually requires an optimization process even for a single target \cite{pytheus} and no general method for discovering the construction rules (python code) is established. An instructive example for this asymmetry is the problem of finding the integral vs. the derivative of a mathematical function. This has been previously explored by \cite{lample2019deep}. As there exist clear rules for differentiation, the authors could generate a large number of random functions and compute their derivatives. They then trained a sequence to sequence transformer to translate in the reverse direction, which is the more difficult task of integration.\\
Similar to the approach by \cite{lample2019deep}, we want to train a sequence to sequence transformer to translate in the hard direction (from quantum states to python code). Because the opposite direction is a straightforward computation, we can produce a large amount of data to train the model by generating random codes (see Fig. \ref{fig:datagen_main}). Our training data consists of two sequences for each sample. The process of translating sequence A (list of states) to sequence B (meta-solution in form of python code) is the difficult direction, which the model is trained to do.\\
Using a simple set of rules, we can generate a random python code, which contains instructions for how to set up an experiment. Each code contains the variable index $N$. This means that the code will result in a different experimental setup for each value of $N$. Simulating the experiment for $N=0,1,2$, we produce three states (see Fig. \ref{fig:datagen_main}). After computing the states, sequence A has the form \texttt{<SOS>[state 1]<SEP>[state 2]<SEP>[state 3]<EOS>} and sequence B is \texttt{<SOS>[python code]<EOS>}. \texttt{<SOS>} and \texttt{<EOS>} are the start-of-sequence and end-of-sequence tokens and \texttt{<SEP>} is a separation token. In the supplementary Figure \ref{fig:datasupp} we show a full example for how a training sample is generated and tokenized. We curate a vocabulary (shared by both sequences) for the tokenization. For further details on the tokenization, see SI section \ref{sec:SItoken}. The maximum length for both sequences during data generation is 640 tokens. We spend about 50,000 CPU hours on generating $56$ million samples.\\
For the model to successfully generalize to unseen targets, it is advised to select the distribution of the synthetic data carefully \cite{linalgtransformer}. A simple example is that a model trained on random samples containing states with three polarizational modes can have difficulties solving a task containing states with only two modes, even though would be expected to be easier, because it is a subspace. To ensure performance on a diverse range of possibly interesting subspaces, we generate separate datasets at different levels of difficulty and specialization (length of states and codes, number of modes, constraints on phase parameters) and combine them to one final training dataset.
\paragraph{Training (learn \texttt{A→B})}
The transformer is an autoregressive model designed to predict a probability distribution over the vocabulary for the next token given all previous tokens. A readily trained model can then be used generate tokens by sampling tokens from the distribution and adding them to the sequence one after the other. We train the model with a standard encoder-decoder transformer architecture \cite{transformer}, with Pre-Layer Normalization \cite{xiong2020layer}. We choose the dimensions $n_{emb} = 512$, $n_{\text{layer}}=18$, $n_{\text{heads}}=8$. We use a learned positional encoding \cite{gehring2017convolutional}, as we are not attempting to apply our model to unseen lengths. The model has approximately $133$ million parameters and is trained from random initialization for 750k steps with a batch size of $256$ (approximately $2.5$ epochs on a dataset of $56$ million samples). The learning rate of the Adam optimizer \cite{kingma2014adam} was $10^{-4}$ for the first epoch and was then lowered to $10^{-5}$. The training was performed on four A100-40GB GPUs. 
\begin{figure}[ht!]
    \centering
    \includegraphics[width = 0.493\textwidth]{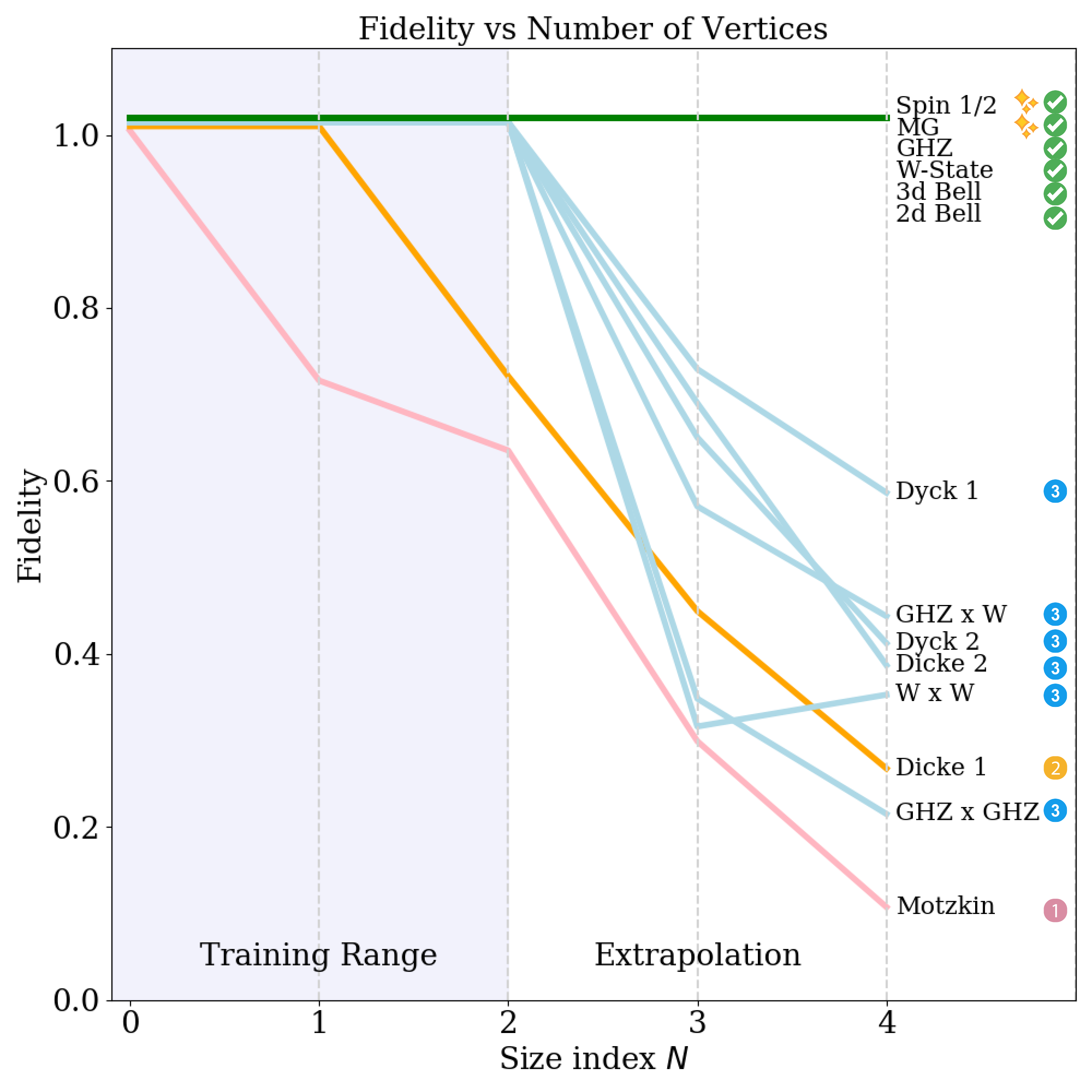}
    \caption{\textbf{Our approach discovers two previously unknown and four previously known generalizations.} We show the resulting fidelities of the best produced code for 14 of the 20 target classes. The fidelity ranges from $0$ (orthogonal to target) to $1$ (perfect match). The green line represents the six target classes which our approach produces codes which correctly extrapolate beyond the first three elements. The blue lines show classes for which the best generated codes have fidelity one for the first three elements of the class, but do not extrapolate beyond. These cases are interesting as the model is still successful in generating a code which matches the three states provided as an input sequence, but the output for $N\geq3$ does not match what we expect. The orange and red line are representatives of the $8$ cases, for which the model was not able to predict correct solutions up to $N=3$. The full table of target classes with their maximum correct $N$ is shown in the supplement. The supplement also contains extended version of this plot for all twenty classes and for values of $N$ up to 7 (Fig. \ref{fig:fidsvsvert_extended}).}
    \label{fig:fidvsvert}
\end{figure}
\begin{figure*}[ht!]
    \centering
    \includegraphics[width = 0.9\textwidth]{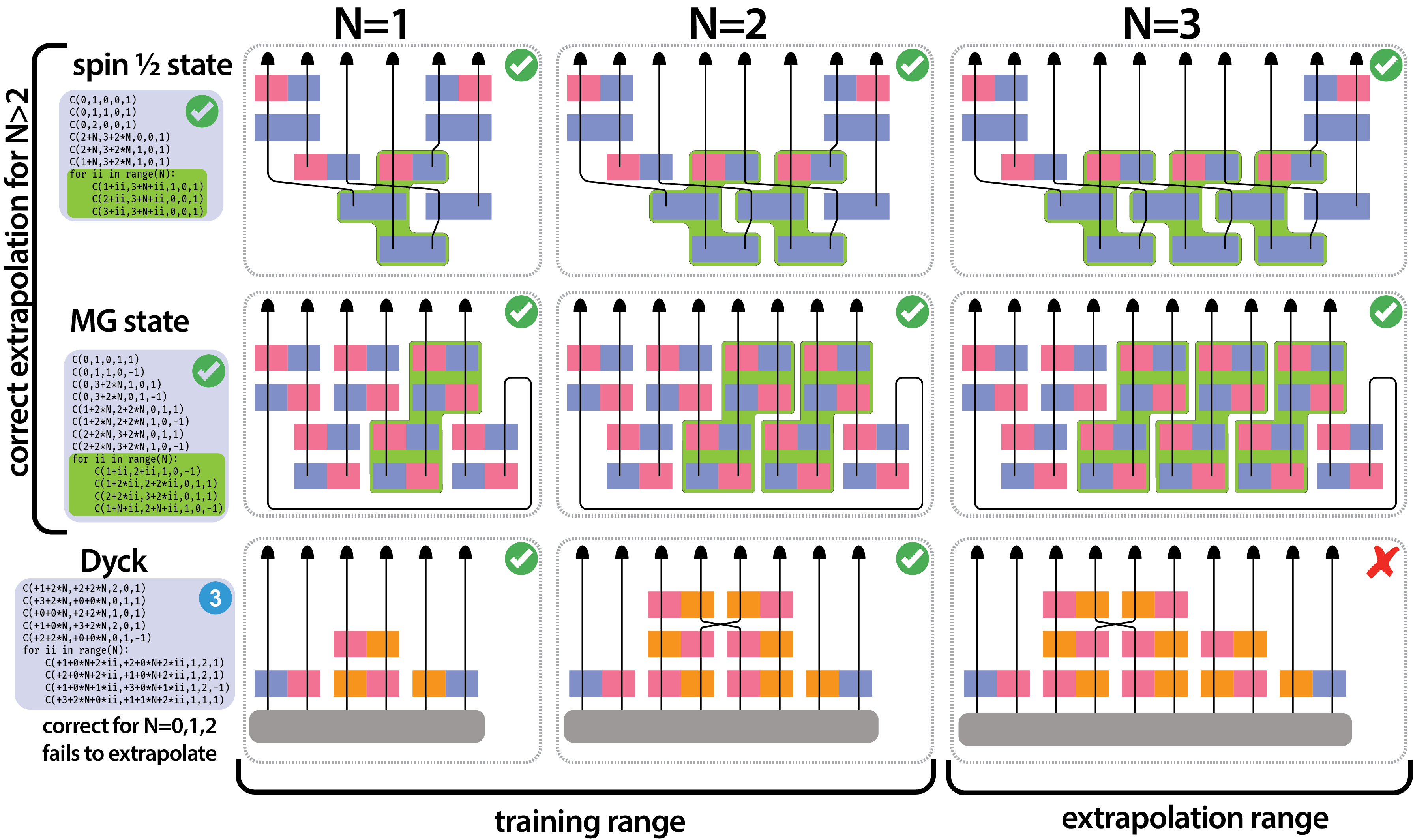}
    \caption{\textbf{Experimental setups for previously unknown solutions exhibit comprehensible patterns.} In the two top rows we show two previously unknown constructions discovered by our approach. For the spin$\frac{1}{2}$ states and the Majumdar-Ghosh states (described in more detail in \cite{pytheus}). For each of the two examples, the code produces the correct experimental setup for the three states used to prompt the model but also for higher particle numbers, indicating that the model was able to pick up on the pattern and write a correct code for the entire class of states. We highlight in green the 'building blocks', which are repeated multiple times as the particle number grows (stemming from lines written in the for loop). The bottom row shows a code for the Dyck 1 state. The setups generated by this code produce the correct state up to the third iteration, but are missing terms for indices $N>2$. This means that the model was able to solve the task it was trained to do (match the first three states), but failed at the meta task of picking up on the pattern we intended it to match beyond the first three examples. It is also notable that in contrast to the other two examples, all setups produced for the Dyck 1 state also contained additional crystals which did not actually contribute to the resulting quantum state. We have omitted them by covering them by a grey rounded rectangle.}
    \label{fig:setups}
\end{figure*}
\subsection*{Results}
\paragraph{Application to selected quantum state classes without known solutions}
Our goal is now to apply the trained model to targets for which the code (sequence B) is unknown. Random generated data is abundant and is thus useful to train our model, but our aim is to discover codes for quantum state classes of particular interest (because of particular mathematical or physical properties). We have compiled a collection of twenty target classes based on a collection of quantum states found in \cite{pytheus} -- all of these states have exceptional properties that have been studied previously, for example in the context of quantum simulations or quantum communication. The first three states of each target class are explicitly shown in the appendix. They are expressed as strings in the same way in which they are given to the model as input.\\
For four out of the 20 targets, meta-solutions were hand-crafted by researchers in the past. For 16 of the 20 target classes, no meta-solution was known before our work.  Furthermore, we do not even know whether a solution can exist at all with the quantum-physical resources we provide (e.g. number of particles necessary to realize a state, and amount of quantum entanglement). Thus, every meta-solution from these 16 states is not a \textit{re}discovery, but a genuine unbiased discovery.\\
\paragraph{Hypothesis generation}
We let the model produce hypotheses for possible meta-solutions by giving the first three states of a target class as an input sequence. For each target, we can generate many varying predictions by picking randomly from the most likely next tokens with top-$p$ sampling ($p=0.5$ and temperature $0.2$, \cite{chen2021evaluating, li2023starcoder}) for four hours on one RTX 6000 GPU, which produces 800-2500 samples (depending on the target class) at a speed of $\sim25$ tokens per second. We evaluate the resulting codes by executing them to produce experimental setups for $N = 0,1,2,3,4$ (training data was generated only for $N = 0,1,2$). We compute the states which are produced by these setups and compute their fidelity with respect to the corresponding target state. The fidelity ranges from $0$ (orthogonal to target) to $1$ (perfect match). In Fig. \ref{fig:fidvsvert} we show the fidelities of the best sample for fourteen of the twenty target classes.\\
Here, the majority of sampling time is spent on model inference, while computing the fidelities for $N = 0,1,2,3,4$ is relatively fast. Evaluating fidelities for much larger scales (such as $N=20$) can take significantly more time. For this reason, we only include $N=3,4$ as a an initial test for correct extrapolation. Successful samples are then evaluated for higher $N$ as well. In the supplementary figures \ref{fig:hist} and \ref{fig:rand_vs_pred} we show the distribution of fidelities for all generated samples and how they compare to the training data. We see that for the successful classes more than 1 percent of produced codes are correct, which means that a shorter sampling time would have sufficed.\\
We have found that our final trained model produces exclusively valid codes for all target classes at low temperatures, even if they do not generate the correct states. This is not obvious, as there are many ways in which a randomly generated, syntactically correct code can still be invalid.
\paragraph{Successful meta-design of codes (6 out of 20 cases)} Before training the model we prepared a set of 20 classes of quantum states as targets for our method. The target classes we consider here all have wave functions that can be expressed as a function $\ket{\psi(N)}$ of a positive integer $N$. We require the number of particles in $\ket{\psi(N)}$ to be less than or equal to $2N+4$ as this is the maximum system size which we allow for during data generation.\\
In Fig. \ref{fig:fidvsvert} we show the fidelities of the best sample for fourteen of the twenty target classes. The best sample is chosen by filtering for samples with the highest $N$ such that all fidelities up to order $N$ are equal to one and then choosing the one with the highest average fidelity for all $N\leq 4$. We find six target classes that our model can solve perfectly. For these classes, the output extrapolates beyond what the model was trained to do, i.e. match the states for $N=0,1,2$.\\
For four famous classes of quantum states (GHZ, W 2d-Bell and 3d-Bell), we knew that there exists a construction rule for experiments with $2N+4$ particles for arbitrary $N$, which act as a baseline check for the capability of our method. Our model rediscovered all four meta-solutions of these states.\\
Most importantly, two out of six classes which our method successfully solves, were previously unknown and thus constitute a genuine discovery. The first previously unknown case is the general Spin-$\frac{1}{2}$ state. There, no two neighboring spin-ups appear in the ground state. In Rydberg-atom experiment, this situation occurs due to the Rydberg blockade \cite{pytheus,bernien2017probing}, however so far it was unknown how to build such an entangled class for photonic systems. The second novel class contains the states of the famous Majumdar-Ghosh Model in condensed matter physics . In this one-dimensional Heisenberg chain, the value of the next-nearest-neighbor interaction is half the value of the nearest-neighbor antiferromagnetic exchange interaction \cite{pytheus,chhajlany2007entanglement}.\\
\paragraph{Codes with unexpected generalizations (6 out of 20 cases)}
For these cases the model produces codes, which generate the correct states for the first three elements of the class, but produces experiments that produce states other than the expected ones. These cases are interesting to examine because the model successfully performs the task it was trained for, as the first three states match the input sequence. The fact that it does not continue to match the target beyond $N=3$ is due to a degree of ambiguity that exists for the continuation of any infinite sequence if only a finite number of elements is given. A possible way to narrow (but not remove) this ambiguity in our application would be to train the model on more than three elements, which has to be traded off with an additional computational cost due to a longer sequence length. Further, the output is highly influenced by the synthetic data. The model will be more likely to produce an output which closer fits the distribution of data it was trained on. One example, the Dyck 1 states, is shown and analyzed in Fig. \ref{fig:setups}. This is an example which does not follow the intended pattern for $N\geq 3$, but produces valid states regardless, which randomly generated experimental setups generally do not do. There is potential in examining these cases in more detail to see if the pattern they follow is interesting from a physics side, as they might represent new unexplored classes of quantum states.\\
\paragraph{Codes which fail to match the first three states (8 out of 20 cases)}
Four classes match the first two states of the input. Another four classes only match the first state of the input states. There were no examples where the model could not match any input states. These also include cases for which the setups generated by the output code do not produce a valid quantum state at higher indices $N$. These cases could be either too complex for the model to give the correct prediction, or generalisations cannot exist at all for physical reasons, given the experimental resources we provide (e.g. total size of setup). We have found after analyzing our results and testing with direct optimization algorithms that, for example, the GHZ3d x GHZ3d state is not achievable in the six-particle case without additional ancillary photons, meaning that this would not be a solvable task, no matter how well the model is trained.\\

\subsection*{Additional Application to Design of Quantum Circuits and Quantum Graph States}
To show that meta-design can be useful in other fields, we also applied it to finding codes for circuit and graph state design, which can be implemented on existing quantum computing platforms.

\begin{figure*}[ht!]
    \centering
    \includegraphics[width = 0.97\textwidth]{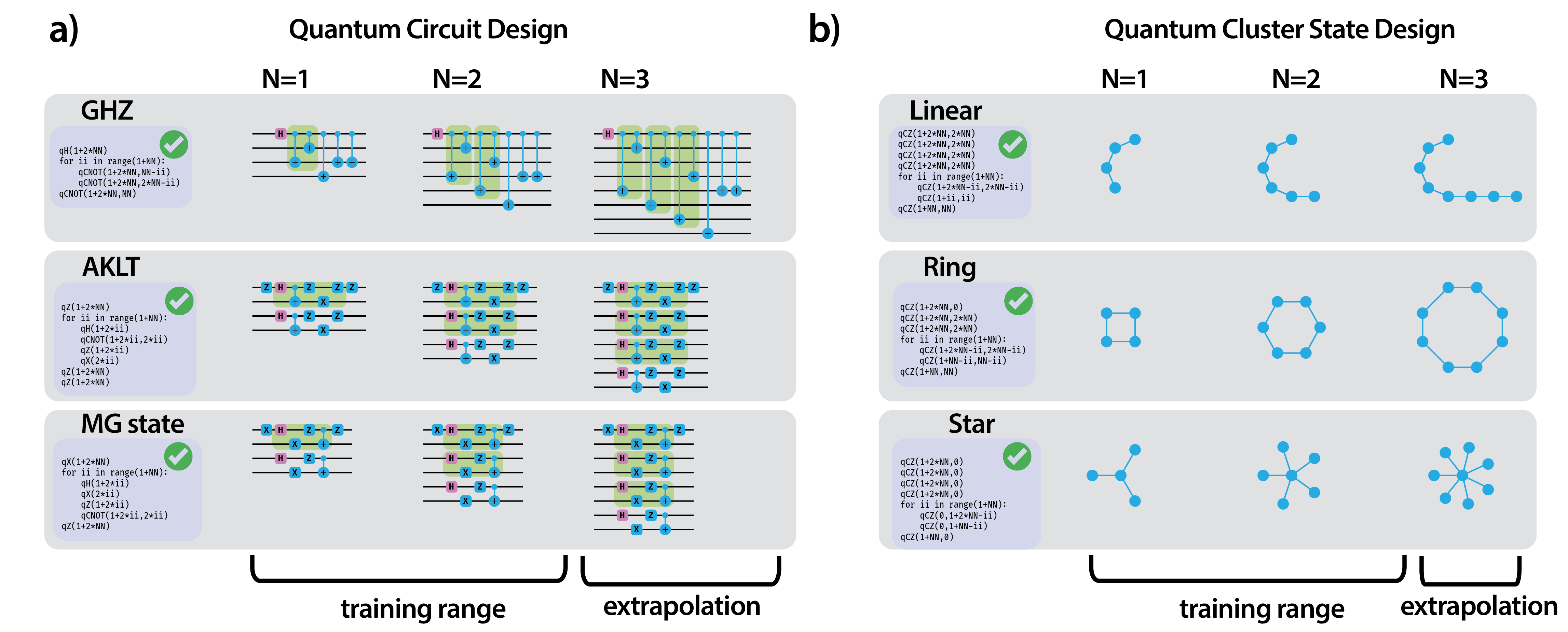}
    \caption{\textbf{Results for additional tasks (quantum circuits and quantum graph states)} a) shows the output codes (blue boxes on the left) of a model trained for quantum cicuit design. We draw the circuit diagram for $N=1,2,3$. Computation of the resulting states confirms that each of the codes (GHZ, AKLT, MG) correctly produces the corresponding target states, even for $N=3$, which shows that the code extrapolates beyond the scales that are seen during training ($N=0,1,2$). b) shows the output codes (blue boxes on the left) of a model trained for quantum graph state. We draw the generated graphs for $N=1,2,3$. Computation of the resulting states confirms that each of the codes (Linear, Ring, Star) correctly produces the corresponding target states, even for $N=3$, which shows that the code extrapolates beyond the scales that are seen during training ($N=0,1,2$).}
    \label{fig:circuit}
\end{figure*}

\paragraph{Quantum Circuits}

Quantum circuits diagrams are an abstract representation of qubits and the transformations that are applied to them in the form of quantum gates, reminiscent of logic gates in classical computing \cite{nielsen2010quantum}. These transformations are unitary and can act on one qubit (e.g. Pauli gates or Hadamard gate), two qubits (e.g. CNOT or CZ), or three qubits (e.g. Toffoli gate or Fredkin gate).

We can also express quantum circuits as code with functions corresponding to specific gates and arguments corresponding to the qubits the gate is applied to (see blue boxes of Fig. \ref{fig:circuit}).

We tested the trained model on the state classes from the main task and successfully discover codes for some of them (Bell, GHZ, AKLT and Majumdar-Ghosh). We visualize some of the solutions in Fig. \ref{fig:circuit}.

\paragraph{Quantum Graph States}
Graph states are a type of entangled quantum states that serve as a universal resource for measurement-based quantum computation (MBQC) \cite{briegel2009measurement}.  Graph states are usually represented as graph states with a regular lattice structure, where each node corresponds to a qubit initialized in the superposition $\frac{1}{\sqrt{2}}(\ket{0}+\ket{1})$ and entangled with its neighbors through controlled-Z (CZ) gates.

We can express the codes for generating classes of graph states by using the same syntax as quantum circuit design and restricting possible gates to the CZ gate.

We show that the model can write code for the three well-known classes of graph states, \textit{linear}, \textit{ring}, and \textit{star}, as shown in Fig. \ref{fig:circuit}.

\paragraph{Training}
We train two separate models for both tasks (A and B). The training procedure was almost the same for both. We generate 9.6M training samples according to rules specified in the supplement [added below]. For these additional tasks we chose slightly smaller models than in the main task. Each model has 12 layers, 8 attention heads and an embedding dimension of 512, resulting in 89M total parameters. Due to the longer sequence length for task B (see supplement), we had to choose different batch sizes (task A: 96, task B: 32). Each model was trained for three days on one node of eight V100 GPUs with AdamW and a learning rate $2\times 10^{-4}$.

\subsection*{Discussion}
We demonstrate how a language model can  create
human-readable Python code (meta-solutions), which solves an entire class of physical design tasks. We discover previously unknown generalizations of experimental setups for quantum state classes of particular interest. The ability to automatically create generalizations is not only a way of generating and extraction of scientific insight, but also offers a decisive advantage over conventional AI-driven design in terms of computational costs.

There are a number of interesting open questions for future research. Firstly, in many design questions, objects can be continuously parametrized. In our work, we constrain ourselves to a subset of special numbers (e.g. integers or $\sqrt(2)$). One can construct arbitrary number approximations by accessing individual integers as tokens, however this process is highly inefficient. It would be interesting to develop a method that can automatically create arbitrary numbers. Secondly, our approach works when large amount of code examples for training can be generated efficiently. As in our examples, the creation of examples was reasonably fast, we did not need to optimize the data efficiency of the models. For other questions, examples might be more computationally expensive, and it will be interesting how to optimize the training data efficiency in such cases. Third, at the moment it is not well understood how the neural scaling law -- an empirical law that relates increased training examples,  modes sizes and compute times with improved values of the loss value for text-generating LLMs \cite{kaplan2020scaling} -- translates to symbolic and mathematical solutions as those in \cite{lample2019deep,cai2024transforming} and ours. Understanding this relation better would help to understand how to improve the model quality, thus reduce the sampling size required to find solutions.

Fourth, and more generally, our method is not constrained to quantum physics but can be directly implemented in other domains, such as the discovery of new microscopes \cite{rodriguez2023xlumina}, new gravitational wave detectors \cite{krenn2023digital}, new experimental hardware for high-energy physics \cite{baydin2021toward}, or the design of new functional molecules \cite{pollice2021data}.\\
At a more abstract level, we see that the application of a powerful intermediate language that can be written and read by both machines and humans can significantly enhance the understandability and generalizability of AI-driven discoveries.




\section*{Data and code availability}
Data for the knowledge graph is accessible on \href{https://zenodo.org/records/14899993}{Zenodo} \cite{metadesignzenodo}. Codes and evaluation data for this work are available on \href{https://github.com/artificial-scientist-lab/meta-design}{Github} \cite{arlt2024meta}.

\section*{Acknowledgements}
The authors thank Ben Newman for useful discussions. The authors also thank Francois Charton for sharing valuable insights on the capabilities and current limitations of language models for mathematics. M.K. acknowledges support by the European Research Council (ERC) under the European Union’s Horizon Europe research and innovation programme (ERC-2024-STG, 101165179, ArtDisQ), and by the German Research Foundation DFG (EXC 2064/1, Project 390727645).



\bibliography{main}

\begin{thebibliography}{69}
\providecommand{\natexlab}[1]{#1}
\providecommand{\url}[1]{\texttt{#1}}
\expandafter\ifx\csname urlstyle\endcsname\relax
  \providecommand{\doi}[1]{doi: #1}\else
  \providecommand{\doi}{doi: \begingroup \urlstyle{rm}\Url}\fi

\bibitem[Alfarano et~al.(2023)Alfarano, Charton, Hayat, and des Ponts~Paristech]{alfarano2023discovering}
Alfarano, A., Charton, F., Hayat, A., and des Ponts~Paristech, C.-E.
\newblock Discovering lyapunov functions with transformers.
\newblock In \emph{The 3rd Workshop on Mathematical Reasoning and AI at NeurIPS'23}, 2023.

\bibitem[Alnuqaydan et~al.()Alnuqaydan, Gleyzer, Prosper, Reinhardt, Anand, and Charton]{alnuqaydansymbolic}
Alnuqaydan, A., Gleyzer, S., Prosper, H.~B., Reinhardt, E.~A., Anand, N., and Charton, F.
\newblock Symbolic machine learning for high energy physics calculations.

\bibitem[Arlt(2025)]{metadesignzenodo}
Arlt, S.
\newblock Model checkpoints and example training data for ``{Meta-Designing} quantum experiments with language models'', 2025.

\bibitem[Arlt et~al.(2024)Arlt, Duan, Li, Xie, Wu, and Krenn]{arlt2024meta}
Arlt, S., Duan, H., Li, F., Xie, S.~M., Wu, Y., and Krenn, M.
\newblock Meta-designing quantum experiments with language models.
\newblock \emph{arXiv preprint arXiv:2406.02470}, 2024.
\newblock \doi{https://doi.org/10.48550/arXiv.2406.02470}.

\bibitem[Aslam et~al.(2023)Aslam, Zhou, Urbach, Turner, Walsworth, Lukin, and Park]{Aslam2023}
Aslam, N., Zhou, H., Urbach, E.~K., Turner, M.~J., Walsworth, R.~L., Lukin, M.~D., and Park, H.
\newblock Quantum sensors for biomedical applications.
\newblock \emph{Nature Reviews Physics}, 5\penalty0 (3):\penalty0 157–169, February 2023.

\bibitem[Aspuru-Guzik \& Walther(2012)Aspuru-Guzik and Walther]{AspuruGuzik2012}
Aspuru-Guzik, A. and Walther, P.
\newblock Photonic quantum simulators.
\newblock \emph{Nature physics}, 8\penalty0 (4):\penalty0 285--291, 2012.

\bibitem[Austin et~al.(2021)Austin, Odena, Nye, Bosma, Michalewski, Dohan, Jiang, Cai, Terry, Le, et~al.]{austin2021program}
Austin, J., Odena, A., Nye, M., Bosma, M., Michalewski, H., Dohan, D., Jiang, E., Cai, C., Terry, M., Le, Q., et~al.
\newblock Program synthesis with large language models.
\newblock \emph{arXiv preprint arXiv:2108.07732}, 2021.

\bibitem[Barman et~al.(2024)Barman, Caron, Claassen, and De~Regt]{barman2024towards}
Barman, K.~G., Caron, S., Claassen, T., and De~Regt, H.
\newblock Towards a benchmark for scientific understanding in humans and machines.
\newblock \emph{Minds and Machines}, 34\penalty0 (1):\penalty0 1--16, 2024.

\bibitem[Baydin et~al.(2021)Baydin, Cranmer, de~Castro~Manzano, Delaere, Derkach, Donini, Dorigo, Giammanco, Kieseler, Layer, Louppe, Ratnikov, Strong, Tosi, Ustyuzhanin, Vischia, and Yarar]{baydin2021toward}
Baydin, A.~G., Cranmer, K., de~Castro~Manzano, P., Delaere, C., Derkach, D., Donini, J., Dorigo, T., Giammanco, A., Kieseler, J., Layer, L., Louppe, G., Ratnikov, F., Strong, G., Tosi, M., Ustyuzhanin, A., Vischia, P., and Yarar, H.
\newblock Toward machine learning optimization of experimental design.
\newblock \emph{Nuclear Physics News}, 31\penalty0 (1):\penalty0 25--28, 2021.

\bibitem[Bernien et~al.(2017)Bernien, Schwartz, Keesling, Levine, Omran, Pichler, Choi, Zibrov, Endres, Greiner, et~al.]{bernien2017probing}
Bernien, H., Schwartz, S., Keesling, A., Levine, H., Omran, A., Pichler, H., Choi, S., Zibrov, A.~S., Endres, M., Greiner, M., et~al.
\newblock Probing many-body dynamics on a 51-atom quantum simulator.
\newblock \emph{Nature}, 551\penalty0 (7682):\penalty0 579--584, 2017.

\bibitem[Briegel et~al.(2009)Briegel, Browne, D{\"u}r, Raussendorf, and Van~den Nest]{briegel2009measurement}
Briegel, H.~J., Browne, D.~E., D{\"u}r, W., Raussendorf, R., and Van~den Nest, M.
\newblock Measurement-based quantum computation.
\newblock \emph{Nature Physics}, 5\penalty0 (1):\penalty0 19--26, 2009.

\bibitem[Cai et~al.(2024)Cai, Merz, Charton, Nolte, Wilhelm, Cranmer, and Dixon]{cai2024transforming}
Cai, T., Merz, G.~W., Charton, F., Nolte, N., Wilhelm, M., Cranmer, K., and Dixon, L.~J.
\newblock Transforming the bootstrap: using transformers to compute scattering amplitudes in planar n = 4 super yang--mills theory.
\newblock \emph{Machine Learning: Science and Technology}, 5\penalty0 (3):\penalty0 035073, 2024.

\bibitem[Charton(2021)]{linalgtransformer}
Charton, F.
\newblock Linear algebra with transformers.
\newblock \emph{arXiv:2112.01898}, 2021.

\bibitem[Chen et~al.(2021)Chen, Tworek, Jun, Yuan, Pinto, Kaplan, Edwards, Burda, Joseph, Brockman, Ray, Puri, Krueger, Petrov, Khlaaf, Sastry, Mishkin, Chan, Gray, Ryder, Pavlov, Power, Kaiser, Bavarian, Winter, Tillet, Petroski~Such, Cummings, Plappert, Chantzis, Barnes, Herbert-Voss, Guss, Nichol, Paino, Tezak, Tang, Babuschkin, Balaji, Jain, Saunders, Hesse, Carr, Leike, Achiam, Misra, Morikawa, Radford, Knight, Brundage, Murati, Mayer, Welinder, McGrew, Amodei, McCandlish, Sutskever, and Zaremba]{chen2021evaluating}
Chen, M., Tworek, J., Jun, H., Yuan, Q., Pinto, H. P. d.~O., Kaplan, J., Edwards, H., Burda, Y., Joseph, N., Brockman, G., Ray, A., Puri, R., Krueger, G., Petrov, M., Khlaaf, H., Sastry, G., Mishkin, P., Chan, B., Gray, S., Ryder, N., Pavlov, M., Power, A., Kaiser, L., Bavarian, M., Winter, C., Tillet, P., Petroski~Such, F., Cummings, D., Plappert, M., Chantzis, F., Barnes, E., Herbert-Voss, A., Guss, W.~H., Nichol, A., Paino, A., Tezak, N., Tang, J., Babuschkin, I., Balaji, S., Jain, S., Saunders, W., Hesse, C., Carr, A.~N., Leike, J., Achiam, J., Misra, V., Morikawa, E., Radford, A., Knight, M., Brundage, M., Murati, M., Mayer, K., Welinder, P., McGrew, B., Amodei, D., McCandlish, S., Sutskever, I., and Zaremba, W.
\newblock Evaluating large language models trained on code.
\newblock \emph{arXiv:2107.03374}, 2021.

\bibitem[Chhajlany et~al.(2007)Chhajlany, Tomczak, W{\'o}jcik, and Richter]{chhajlany2007entanglement}
Chhajlany, R.~W., Tomczak, P., W{\'o}jcik, A., and Richter, J.
\newblock Entanglement in the majumdar-ghosh model.
\newblock \emph{Physical Review A}, 75\penalty0 (3):\penalty0 032340, 2007.

\bibitem[Cornish et~al.(2024)Cornish, Tarbutt, and Hazzard]{Cornish2024-mg}
Cornish, S.~L., Tarbutt, M.~R., and Hazzard, K. R.~A.
\newblock Quantum computation and quantum simulation with ultracold molecules.
\newblock \emph{Nature Physics}, 20\penalty0 (5):\penalty0 730--740, 2024.

\bibitem[Couteau et~al.(2023)Couteau, Barz, Durt, Gerrits, Huwer, Prevedel, Rarity, Shields, and Weihs]{Couteau2023-ql}
Couteau, C., Barz, S., Durt, T., Gerrits, T., Huwer, J., Prevedel, R., Rarity, J., Shields, A., and Weihs, G.
\newblock Applications of single photons to quantum communication and computing.
\newblock \emph{Nature Reviews Physics}, 5\penalty0 (6):\penalty0 326--338, 2023.

\bibitem[De~Regt(2017)]{de2017understanding}
De~Regt, H.~W.
\newblock \emph{Understanding scientific understanding}.
\newblock Oxford University Press, 2017.

\bibitem[DeMille et~al.(2024)DeMille, Hutzler, Rey, and Zelevinsky]{DeMille2024}
DeMille, D., Hutzler, N.~R., Rey, A.~M., and Zelevinsky, T.
\newblock Quantum sensing and metrology for fundamental physics with molecules.
\newblock \emph{Nature Physics}, 20\penalty0 (5):\penalty0 741–749, May 2024.

\bibitem[Dettmers et~al.(2023)Dettmers, Pagnoni, Holtzman, and Zettlemoyer]{dettmers2023}
Dettmers, T., Pagnoni, A., Holtzman, A., and Zettlemoyer, L.
\newblock Qlora: Efficient finetuning of quantized llms.
\newblock \emph{Advances in neural information processing systems}, 36:\penalty0 10088--10115, 2023.

\bibitem[Dou et~al.(2024)Dou, Liu, Jia, Xiong, Zhou, Shen, Shan, Huang, Wang, Fan, et~al.]{dou2024stepcoder}
Dou, S., Liu, Y., Jia, H., Xiong, L., Zhou, E., Shen, W., Shan, J., Huang, C., Wang, X., Fan, X., et~al.
\newblock Stepcoder: Improve code generation with reinforcement learning from compiler feedback.
\newblock \emph{arXiv preprint arXiv:2402.01391}, 2024.

\bibitem[Flamini et~al.(2018)Flamini, Spagnolo, and Sciarrino]{Flamini2018}
Flamini, F., Spagnolo, N., and Sciarrino, F.
\newblock Photonic quantum information processing: a review.
\newblock \emph{Reports on Progress in Physics}, 82\penalty0 (1):\penalty0 016001, 2018.

\bibitem[Gedeon et~al.(2023)Gedeon, Hassan, and Lesina]{gedeon2023free}
Gedeon, J., Hassan, E., and Lesina, A.~C.
\newblock Free-form inverse design of arbitrary dispersive materials in nanophotonics.
\newblock \emph{arXiv:2305.00234}, 2023.

\bibitem[Gehring et~al.(2017)Gehring, Auli, Grangier, Yarats, and Dauphin]{gehring2017convolutional}
Gehring, J., Auli, M., Grangier, D., Yarats, D., and Dauphin, Y.~N.
\newblock Convolutional sequence to sequence learning.
\newblock In \emph{International conference on machine learning}, pp.\  1243--1252. PMLR, 2017.

\bibitem[Goel et~al.(2024)Goel, Leedumrongwatthanakun, Valencia, McCutcheon, Tavakoli, Conti, Pinkse, and Malik]{Goel2024}
Goel, S., Leedumrongwatthanakun, S., Valencia, N.~H., McCutcheon, W., Tavakoli, A., Conti, C., Pinkse, P.~W., and Malik, M.
\newblock Inverse design of high-dimensional quantum optical circuits in a complex medium.
\newblock \emph{Nature Physics}, pp.\  1--8, 2024.

\bibitem[Greenberger et~al.(1990)Greenberger, Horne, Shimony, and Zeilinger]{Greenberger1990}
Greenberger, D.~M., Horne, M.~A., Shimony, A., and Zeilinger, A.
\newblock Bell’s theorem without inequalities.
\newblock \emph{American Journal of Physics}, 58\penalty0 (12):\penalty0 1131–1143, December 1990.

\bibitem[Javadi-Abhari et~al.(2024)Javadi-Abhari, Treinish, Krsulich, Wood, Lishman, Gacon, Martiel, Nation, Bishop, Cross, et~al.]{javadi2024quantum}
Javadi-Abhari, A., Treinish, M., Krsulich, K., Wood, C.~J., Lishman, J., Gacon, J., Martiel, S., Nation, P.~D., Bishop, L.~S., Cross, A.~W., et~al.
\newblock Quantum computing with qiskit.
\newblock \emph{arXiv:2405.08810}, 2024.

\bibitem[Kamienny et~al.(2022)Kamienny, d\textquotesingle Ascoli, Lample, and Charton]{charton22}
Kamienny, P.-a., d\textquotesingle Ascoli, S., Lample, G., and Charton, F.
\newblock End-to-end symbolic regression with transformers.
\newblock In Koyejo, S., Mohamed, S., Agarwal, A., Belgrave, D., Cho, K., and Oh, A. (eds.), \emph{Advances in Neural Information Processing Systems}, volume~35, pp.\  10269--10281. Curran Associates, Inc., 2022.

\bibitem[Kaplan et~al.(2020)Kaplan, McCandlish, Henighan, Brown, Chess, Child, Gray, Radford, Wu, and Amodei]{kaplan2020scaling}
Kaplan, J., McCandlish, S., Henighan, T., Brown, T.~B., Chess, B., Child, R., Gray, S., Radford, A., Wu, J., and Amodei, D.
\newblock Scaling laws for neural language models.
\newblock \emph{arXiv:2001.08361}, 2020.

\bibitem[Kingma \& Ba(2014)Kingma and Ba]{kingma2014adam}
Kingma, D.~P. and Ba, J.
\newblock Adam: A method for stochastic optimization.
\newblock \emph{arXiv preprint arXiv:1412.6980}, 2014.

\bibitem[Knott(2016)]{Knott2016}
Knott, P.
\newblock A search algorithm for quantum state engineering and metrology.
\newblock \emph{New Journal of Physics}, 18\penalty0 (7):\penalty0 073033, 2016.

\bibitem[Kottmann(2023)]{Kottmann2023}
Kottmann, J.~S.
\newblock Molecular quantum circuit design: A graph-based approach.
\newblock \emph{Quantum}, 7:\penalty0 1073, August 2023.

\bibitem[Krenn et~al.(2016)Krenn, Malik, Fickler, Lapkiewicz, and Zeilinger]{krenn2016automated}
Krenn, M., Malik, M., Fickler, R., Lapkiewicz, R., and Zeilinger, A.
\newblock Automated search for new quantum experiments.
\newblock \emph{Physical review letters}, 116\penalty0 (9):\penalty0 090405, 2016.

\bibitem[Krenn et~al.(2020)Krenn, Erhard, and Zeilinger]{Krenn2020}
Krenn, M., Erhard, M., and Zeilinger, A.
\newblock Computer-inspired quantum experiments.
\newblock \emph{Nature Reviews Physics}, 2\penalty0 (11):\penalty0 649–661, September 2020.

\bibitem[Krenn et~al.(2021)Krenn, Kottmann, Tischler, and Aspuru-Guzik]{theseus}
Krenn, M., Kottmann, J.~S., Tischler, N., and Aspuru-Guzik, A.
\newblock Conceptual understanding through efficient automated design of quantum optical experiments.
\newblock \emph{Physical Review X}, 11\penalty0 (3):\penalty0 031044, 2021.

\bibitem[Krenn et~al.(2022)Krenn, Pollice, Guo, Aldeghi, Cervera-Lierta, Friederich, dos Passos~Gomes, H{\"a}se, Jinich, Nigam, Yao, and Aspuru-Guzik]{Krenn2022}
Krenn, M., Pollice, R., Guo, S.~Y., Aldeghi, M., Cervera-Lierta, A., Friederich, P., dos Passos~Gomes, G., H{\"a}se, F., Jinich, A., Nigam, A., Yao, Z., and Aspuru-Guzik, A.
\newblock On scientific understanding with artificial intelligence.
\newblock \emph{Nature Reviews Physics}, 4\penalty0 (12):\penalty0 761--769, 2022.

\bibitem[Krenn et~al.(2025)Krenn, Drori, and Adhikari]{krenn2023digital}
Krenn, M., Drori, Y., and Adhikari, R.~X.
\newblock Digital discovery of interferometric gravitational wave detectors.
\newblock \emph{Physical Review X}, 15\penalty0 (2):\penalty0 021012, 2025.

\bibitem[Krzysztof \& Eisenecker(2000)Krzysztof and Eisenecker]{krzysztof2000generative}
Krzysztof, C. and Eisenecker, U.~W.
\newblock \emph{Generative Programming: Methods, Tools and Applications}.
\newblock Addison-Wesley, 2000.

\bibitem[Kviatkovsky et~al.(2020)Kviatkovsky, Chrzanowski, Avery, Bartolomaeus, and Ramelow]{kviatkovsky2020microscopy}
Kviatkovsky, I., Chrzanowski, H.~M., Avery, E.~G., Bartolomaeus, H., and Ramelow, S.
\newblock Microscopy with undetected photons in the mid-infrared.
\newblock \emph{Science Advances}, 6\penalty0 (42):\penalty0 eabd0264, 2020.

\bibitem[Lample \& Charton(2019)Lample and Charton]{lample2019deep}
Lample, G. and Charton, F.
\newblock Deep learning for symbolic mathematics.
\newblock \emph{arXiv:1912.01412}, 2019.

\bibitem[Landgraf et~al.(2025)Landgraf, Peano, and Marquardt]{https://doi.org/10.48550/arxiv.2404.14887}
Landgraf, J., Peano, V., and Marquardt, F.
\newblock Automated discovery of coupled-mode setups.
\newblock \emph{Physical Review X}, 15\penalty0 (2):\penalty0 021038, 2025.

\bibitem[Lemos et~al.(2014)Lemos, Borish, Cole, Ramelow, Lapkiewicz, and Zeilinger]{lemos2014quantum}
Lemos, G.~B., Borish, V., Cole, G.~D., Ramelow, S., Lapkiewicz, R., and Zeilinger, A.
\newblock Quantum imaging with undetected photons.
\newblock \emph{Nature}, 512\penalty0 (7515):\penalty0 409--412, 2014.

\bibitem[Li et~al.(2023)Li, Ben~Allal, Zi, Muennighoff, Kocetkov, Mou, Marone, Akiki, Li, Chim, Liu, Zheltonozhskii, Zhuo, Wang, Dehaene, Davaadorj, Lamy-Poirier, Monteiro, Shliazhko, Gontier, Meade, Zebaze, Yee, Umapathi, Zhu, Lipkin, Oblokulov, Wang, Murthy, Stillerman, Patel, Abulkhanov, Zocca, Dey, Zhang, Fahmy, Bhattacharyya, Yu, Singh, Luccioni, Villegas, Kunakov, Zhdanov, Romero, Lee, Timor, Ding, Schlesinger, Schoelkopf, Ebert, Dao, Mishra, Gu, Robinson, Anderson, Dolan-Gavitt, Contractor, Reddy, Fried, Bahdanau, Jernite, Ferrandis, Hughes, Wolf, Guha, von Werra, and de~Vries]{li2023starcoder}
Li, R., Ben~Allal, L., Zi, Y., Muennighoff, N., Kocetkov, D., Mou, C., Marone, M., Akiki, C., Li, J., Chim, J., Liu, Q., Zheltonozhskii, E., Zhuo, T.~Y., Wang, T., Dehaene, O., Davaadorj, M., Lamy-Poirier, J., Monteiro, J., Shliazhko, O., Gontier, N., Meade, N., Zebaze, A., Yee, M.-H., Umapathi, L.~K., Zhu, J., Lipkin, B., Oblokulov, M., Wang, Z., Murthy, R., Stillerman, J., Patel, S.~S., Abulkhanov, D., Zocca, M., Dey, M., Zhang, Z., Fahmy, N., Bhattacharyya, U., Yu, W., Singh, S., Luccioni, S., Villegas, P., Kunakov, M., Zhdanov, F., Romero, M., Lee, T., Timor, N., Ding, J., Schlesinger, C., Schoelkopf, H., Ebert, J., Dao, T., Mishra, M., Gu, A., Robinson, J., Anderson, C.~J., Dolan-Gavitt, B., Contractor, D., Reddy, S., Fried, D., Bahdanau, D., Jernite, Y., Ferrandis, C.~M., Hughes, S., Wolf, T., Guha, A., von Werra, L., and de~Vries, H.
\newblock Starcoder: may the source be with you!
\newblock \emph{arXiv:2305.06161}, 2023.

\bibitem[Li et~al.(2022)Li, Choi, Chung, Kushman, Schrittwieser, Leblond, Eccles, Keeling, Gimeno, Dal~Lago, et~al.]{li2022competition}
Li, Y., Choi, D., Chung, J., Kushman, N., Schrittwieser, J., Leblond, R., Eccles, T., Keeling, J., Gimeno, F., Dal~Lago, A., et~al.
\newblock Competition-level code generation with alphacode.
\newblock \emph{Science}, 378\penalty0 (6624):\penalty0 1092--1097, 2022.

\bibitem[Ma et~al.(2021)Ma, Liu, Kudyshev, Boltasseva, Cai, and Liu]{Ma2020}
Ma, W., Liu, Z., Kudyshev, Z.~A., Boltasseva, A., Cai, W., and Liu, Y.
\newblock Deep learning for the design of photonic structures.
\newblock \emph{Nature Photonics}, 15\penalty0 (2):\penalty0 77--90, 2021.

\bibitem[MacLellan et~al.(2024)MacLellan, Roztocki, Czischek, and Melko]{maclellan2024end}
MacLellan, B., Roztocki, P., Czischek, S., and Melko, R.~G.
\newblock End-to-end variational quantum sensing.
\newblock \emph{npj Quantum Information}, 10\penalty0 (1):\penalty0 118, 2024.

\bibitem[Madsen et~al.(2022)Madsen, Laudenbach, Askarani, Rortais, Vincent, Bulmer, Miatto, Neuhaus, Helt, Collins, Lita, Gerrits, Nam, Vaidya, Menotti, Dhand, Vernon, Quesada, and Lavoie]{Madsen2022-ht}
Madsen, L.~S., Laudenbach, F., Askarani, M.~F., Rortais, F., Vincent, T., Bulmer, J. F.~F., Miatto, F.~M., Neuhaus, L., Helt, L.~G., Collins, M.~J., Lita, A.~E., Gerrits, T., Nam, S.~W., Vaidya, V.~D., Menotti, M., Dhand, I., Vernon, Z., Quesada, N., and Lavoie, J.
\newblock Quantum computational advantage with a programmable photonic processor.
\newblock \emph{Nature}, 606\penalty0 (7912):\penalty0 75--81, 2022.

\bibitem[Melko \& Carrasquilla(2024)Melko and Carrasquilla]{melko2024language}
Melko, R.~G. and Carrasquilla, J.
\newblock Language models for quantum simulation.
\newblock \emph{Nature Computational Science}, 4:\penalty0 11--18, 2024.

\bibitem[Molesky et~al.(2018)Molesky, Lin, Piggott, Jin, Vuckovi{\'c}, and Rodriguez]{Molesky2018}
Molesky, S., Lin, Z., Piggott, A.~Y., Jin, W., Vuckovi{\'c}, J., and Rodriguez, A.~W.
\newblock Inverse design in nanophotonics.
\newblock \emph{Nature Photonics}, 12\penalty0 (11):\penalty0 659--670, 2018.

\bibitem[Moreau et~al.(2019)Moreau, Toninelli, Gregory, and Padgett]{Moreau2019}
Moreau, P.-A., Toninelli, E., Gregory, T., and Padgett, M.~J.
\newblock Imaging with quantum states of light.
\newblock \emph{Nature Reviews Physics}, 1\penalty0 (6):\penalty0 367--380, 2019.

\bibitem[N{\"a}gele \& Marquardt(2024)N{\"a}gele and Marquardt]{https://doi.org/10.48550/arxiv.2311.18588}
N{\"a}gele, M. and Marquardt, F.
\newblock Optimizing zx-diagrams with deep reinforcement learning.
\newblock \emph{Machine Learning: Science and Technology}, 5\penalty0 (3):\penalty0 035077, 2024.

\bibitem[Nichols et~al.(2019)Nichols, Mineh, Rubio, Matthews, and Knott]{Nichols2019}
Nichols, R., Mineh, L., Rubio, J., Matthews, J.~C., and Knott, P.~A.
\newblock Designing quantum experiments with a genetic algorithm.
\newblock \emph{Quantum Science and Technology}, 4\penalty0 (4):\penalty0 045012, 2019.

\bibitem[Nielsen \& Chuang(2010)Nielsen and Chuang]{nielsen2010quantum}
Nielsen, M.~A. and Chuang, I.~L.
\newblock \emph{Quantum computation and quantum information}.
\newblock Cambridge university press, 2010.

\bibitem[Novikov et~al.(2025)Novikov, V{\~u}, Eisenberger, Dupont, Huang, Wagner, Shirobokov, Kozlovskii, Ruiz, Mehrabian, et~al.]{novikov2025alphaevolve}
Novikov, A., V{\~u}, N., Eisenberger, M., Dupont, E., Huang, P.-S., Wagner, A.~Z., Shirobokov, S., Kozlovskii, B., Ruiz, F.~J., Mehrabian, A., et~al.
\newblock Alphaevolve: A coding agent for scientific and algorithmic discovery.
\newblock \emph{arXiv preprint arXiv:2506.13131}, 2025.

\bibitem[Ostaszewski et~al.(2021)Ostaszewski, Trenkwalder, Masarczyk, Scerri, and Dunjko]{ostaszewski2021reinforcement}
Ostaszewski, M., Trenkwalder, L.~M., Masarczyk, W., Scerri, E., and Dunjko, V.
\newblock Reinforcement learning for optimization of variational quantum circuit architectures.
\newblock \emph{Advances in Neural Information Processing Systems}, 34:\penalty0 18182--18194, 2021.

\bibitem[Pan et~al.(2000)Pan, Bouwmeester, Daniell, Weinfurter, and Zeilinger]{Pan2000}
Pan, J.-W., Bouwmeester, D., Daniell, M., Weinfurter, H., and Zeilinger, A.
\newblock Experimental test of quantum nonlocality in three-photon greenberger–horne–zeilinger entanglement.
\newblock \emph{Nature}, 403\penalty0 (6769):\penalty0 515–519, February 2000.

\bibitem[Pezz\`e et~al.(2018)Pezz\`e, Smerzi, Oberthaler, Schmied, and Treutlein]{dicke_metrology}
Pezz\`e, L., Smerzi, A., Oberthaler, M.~K., Schmied, R., and Treutlein, P.
\newblock Quantum metrology with nonclassical states of atomic ensembles.
\newblock \emph{Rev. Mod. Phys.}, 90, Sep 2018.

\bibitem[Polino et~al.(2020)Polino, Valeri, Spagnolo, and Sciarrino]{Polino2020}
Polino, E., Valeri, M., Spagnolo, N., and Sciarrino, F.
\newblock Photonic quantum metrology.
\newblock \emph{AVS Quantum Science}, 2\penalty0 (2), 2020.

\bibitem[Pollice et~al.(2021)Pollice, dos Passos~Gomes, Aldeghi, Hickman, Krenn, Lavigne, Lindner-D’Addario, Nigam, Ser, Yao, and Aspuru-Guzik]{pollice2021data}
Pollice, R., dos Passos~Gomes, G., Aldeghi, M., Hickman, R.~J., Krenn, M., Lavigne, C., Lindner-D’Addario, M., Nigam, A., Ser, C.~T., Yao, Z., and Aspuru-Guzik, A.
\newblock Data-driven strategies for accelerated materials design.
\newblock \emph{Accounts of Chemical Research}, 54\penalty0 (4):\penalty0 849--860, 2021.

\bibitem[Prabhu et~al.(2020)Prabhu, Roques-Carmes, Shen, Harris, Jing, Carolan, Hamerly, Baehr-Jones, Hochberg, {\v{C}}eperi{\'c}, et~al.]{prabhu2020accelerating}
Prabhu, M., Roques-Carmes, C., Shen, Y., Harris, N., Jing, L., Carolan, J., Hamerly, R., Baehr-Jones, T., Hochberg, M., {\v{C}}eperi{\'c}, V., et~al.
\newblock Accelerating recurrent ising machines in photonic integrated circuits.
\newblock \emph{Optica}, 7\penalty0 (5):\penalty0 551--558, 2020.

\bibitem[Rodr{\'\i}guez et~al.(2024)Rodr{\'\i}guez, Arlt, M{\"o}ckl, and Krenn]{rodriguez2023xlumina}
Rodr{\'\i}guez, C., Arlt, S., M{\"o}ckl, L., and Krenn, M.
\newblock Automated discovery of experimental designs in super-resolution microscopy with xlumina.
\newblock \emph{Nature Communications}, 15\penalty0 (1):\penalty0 1--14, 2024.

\bibitem[Romera-Paredes et~al.(2024)Romera-Paredes, Barekatain, Novikov, Balog, Kumar, Dupont, Ruiz, Ellenberg, Wang, Fawzi, et~al.]{romera2024mathematical}
Romera-Paredes, B., Barekatain, M., Novikov, A., Balog, M., Kumar, M.~P., Dupont, E., Ruiz, F.~J., Ellenberg, J.~S., Wang, P., Fawzi, O., et~al.
\newblock Mathematical discoveries from program search with large language models.
\newblock \emph{Nature}, 625\penalty0 (7995):\penalty0 468--475, 2024.

\bibitem[Ruiz-Gonzalez et~al.(2023)Ruiz-Gonzalez, Arlt, Petermann, Sayyad, Jaouni, Karimi, Tischler, Gu, and Krenn]{pytheus}
Ruiz-Gonzalez, C., Arlt, S., Petermann, J., Sayyad, S., Jaouni, T., Karimi, E., Tischler, N., Gu, X., and Krenn, M.
\newblock Digital discovery of 100 diverse quantum experiments with pytheus.
\newblock \emph{Quantum}, 7:\penalty0 1204, 2023.

\bibitem[Sapra et~al.(2020)Sapra, Yang, Vercruysse, Leedle, Black, England, Su, Trivedi, Miao, Solgaard, Byer, and Vučković]{sapra2020chip}
Sapra, N.~V., Yang, K.~Y., Vercruysse, D., Leedle, K.~J., Black, D.~S., England, R.~J., Su, L., Trivedi, R., Miao, Y., Solgaard, O., Byer, R.~L., and Vučković, J.
\newblock On-chip integrated laser-driven particle accelerator.
\newblock \emph{Science}, 367\penalty0 (6473):\penalty0 79--83, 2020.

\bibitem[Trinh et~al.(2024)Trinh, Wu, Le, He, and Luong]{alphageometry}
Trinh, T.~H., Wu, Y., Le, Q.~V., He, H., and Luong, T.
\newblock Solving olympiad geometry without human demonstrations.
\newblock \emph{Nature}, 625\penalty0 (7995):\penalty0 476--482, 2024.

\bibitem[Vaswani et~al.(2017)Vaswani, Shazeer, Parmar, Uszkoreit, Jones, Gomez, Kaiser, and Polosukhin]{transformer}
Vaswani, A., Shazeer, N., Parmar, N., Uszkoreit, J., Jones, L., Gomez, A.~N., Kaiser, {\L}., and Polosukhin, I.
\newblock Attention is all you need.
\newblock \emph{Advances in neural information processing systems}, 30, 2017.

\bibitem[Walln{\"o}fer et~al.(2020)Walln{\"o}fer, Melnikov, D{\"u}r, and Briegel]{PRXQuantum.1.010301}
Walln{\"o}fer, J., Melnikov, A.~A., D{\"u}r, W., and Briegel, H.~J.
\newblock Machine learning for long-distance quantum communication.
\newblock \emph{PRX Quantum}, 1\penalty0 (1):\penalty0 010301, 2020.

\bibitem[Xiong et~al.(2020)Xiong, Yang, He, Zheng, Zheng, Xing, Zhang, Lan, Wang, and Liu]{xiong2020layer}
Xiong, R., Yang, Y., He, D., Zheng, K., Zheng, S., Xing, C., Zhang, H., Lan, Y., Wang, L., and Liu, T.
\newblock On layer normalization in the transformer architecture.
\newblock In \emph{International Conference on Machine Learning}, pp.\  10524--10533. PMLR, 2020.

\bibitem[Zen et~al.(2024)Zen, Olle, Colmenarez, Puviani, M{\"u}ller, and Marquardt]{https://doi.org/10.48550/arxiv.2402.17761}
Zen, R., Olle, J., Colmenarez, L., Puviani, M., M{\"u}ller, M., and Marquardt, F.
\newblock Quantum circuit discovery for fault-tolerant logical state preparation with reinforcement learning.
\newblock \emph{arXiv:2402.17761}, 2024.

\end{thebibliography}
\bibliographystyle{icml2024}

\newpage
\appendix
\onecolumn

\section{Solving a class of problems with code - an example from elementary geometry}
Turing-complete programming languages are able to describe any natural process that can be modeled computationally. High-level programming languages like python are also human-readable on top of being executable by computers. This makes them perfect to express general concepts. A very simple example is the following. We can write a python function which describes the action of \textit{walking in a (regular, convex) N-polygon} for an arbitrary number of sides $N\geq 3$.
\begin{verbatim}
def walk_polygon(N):
    for i in range(N):
        forward(1)
        left(360 / N)
\end{verbatim}

This simple program is can express an infinite number of shapes because they belong to a class which follows a pattern.

\section{Details on data generation}\label{sec:SItoken}
\begin{figure}[H]
    \centering
    \includegraphics[width = 0.85\textwidth]{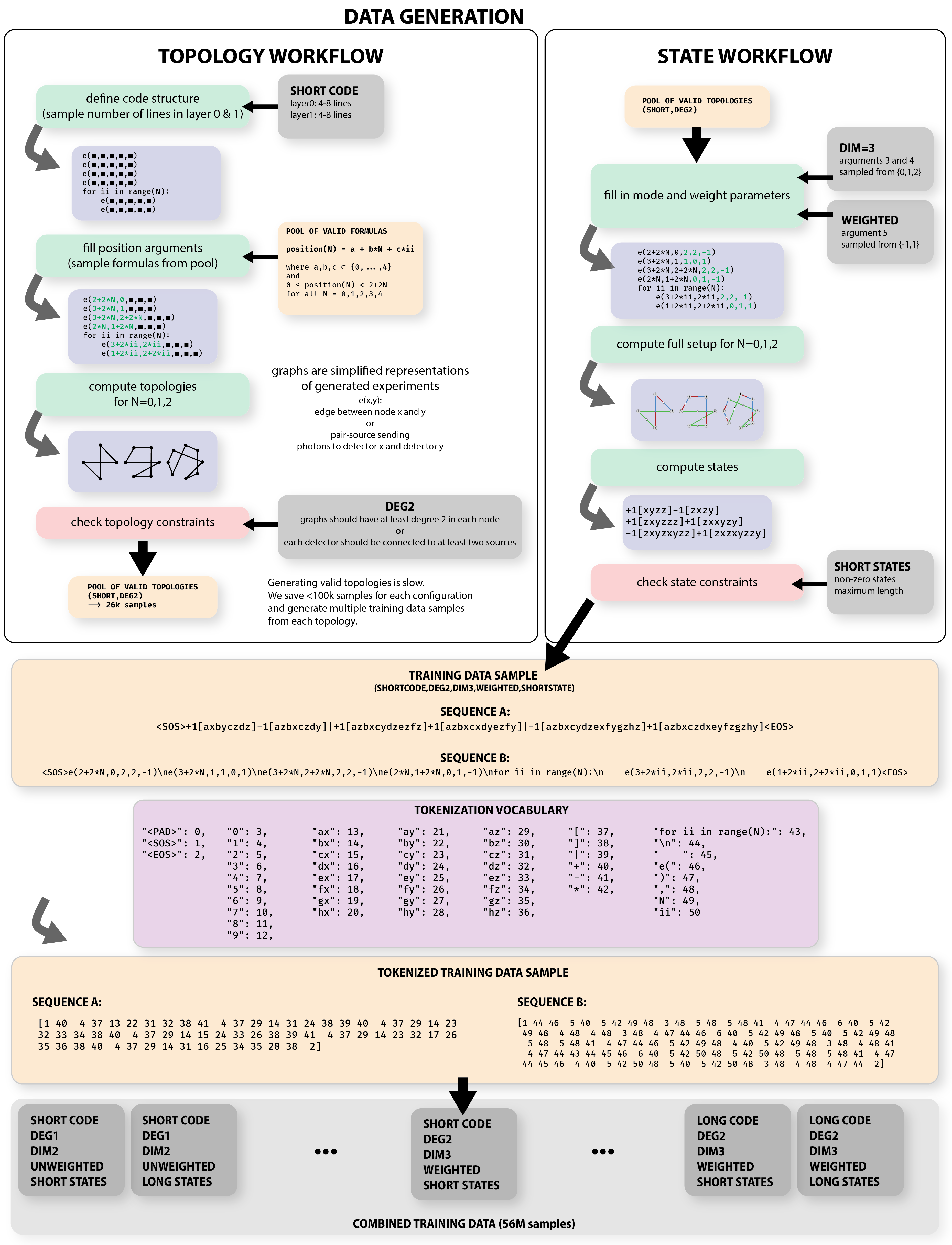}
    \caption{\textbf{Data generation workflow for main task -} the data generation is split into two parts, the \textit{topology workflow (left)} and the \textit{state workflow (right)}. In the topology workflow we find codes which could generate plausible experimental setups. In the state workflow, additional arguments are added to the codes that result in the final states that are created by the experimental setups. The figure follows a real example of a generated sample. The data generation is constrained by a set of hyperparameters (shown in grey boxes). On the bottom we show a final training data sample as it was used for training the model.}
    \label{fig:datasupp}
\end{figure}

The process of data generation is outlined in Fig. \ref{fig:datasupp}. The following subsections describe the hyperparameters that are used to create the full training dataset. The source code (and a pseudocode description) can be found on github.
\subsection*{Hyperparameters for data generation}
The following hyperparameters are to have more direct control on the distribution of generated data. They are not essential to train a working model. One could also train the model on one very weakly constrained data distribution. We chose to combine multiple datasets with different constraints into the training dataset to make sure that e.g. short, two-dimensional states would be well represented in the training data, which would otherwise not be the case in the most general setting.\\
\textbf{Length of code -}All codes follow the structure of $n_0$ lines outside of the a for loop and then $n_1$ lines inside of a for loop. For \textit{long} codes, the numbers are constrained by $4\leq n_0 \leq 12$, $2\leq n_1 \leq 12$. For \textit{short} codes, the numbers are constrained by $4\leq n_0 \leq 8$, $2\leq n_1 \leq 6$\\
\textbf{Minimum Degree of Graphs -}
\textit{DEG1} requires each detector to be connected to at least one pair-source, which is the minimum requirement to achieve a valid output state. \textit{DEG1} requires each detector to be connected to at least two pair-sources, which leads to more advanced entanglement in the resulting quantum state.\\
\textbf{Dimensionality -}
\textit{DIM2} creates quantum states made from two modes (qubits), \textit{DIM3} allows for three quantum modes (qutrits) resulting in more advanced states. We chose to include DIM2 states explicitly because they would be rarely generated by a DIM3 distribution, but there are many potentially interesting quantum states which exist in the DIM2 subspace.\\
\textbf{Weighted/Unweighted -}
\textit{Weighted} allows for the introduction of additional phases in the experimental setups which lead to more advanced interference phenomena. \textit{Unweighted} restricts the setups to not have any phases. Similar to DIM2/DIM3 we chose to include unweighted codes explicitly so this subspace would be better represented in the training data.\\
\textbf{Maximum length of states -}
We restricted the number of terms in the output states in order to limit the length of the sequences that had to be processed by the model and because in our experience more interesting states tend to have fewer terms. We defined \textit{long states} to have a maximum of 8, 16, and 32 terms (for $N=0,1,2$) and \textit{short states} to have a maximum of 6,6,6 terms for ($N=0,1,2$).\\

Final data distribution:\\
Topologies:\\
LONGCODE\_DEG1 - 88k samples\\
LONGCODE\_DEG2 - 79k samples\\
SHORTCODE\_DEG1 - 22k samples\\
SHORTCODE\_DEG2 - 26k samples\\

There are $2^5=32$ combinations of the five hyperparameters, resulting in 32 individual datasets with 1.75M samples each. The final training dataset is generated by combining and shuffling all $32\cdot1.75\cdot10^6=56\cdot10^6$ samples.

\pagebreak

\textbf{Vocabulary -} We constructed a vocabulary specifically for our task as the set of characters used in this task is relatively small. An alternative would be to train a tokenizer using [...] or use pre-trained tokenizers such as [...]. Another possibility would be the simplest possible option of associating each character with one token. We do not see a reason to believe that any of these tokenization strategies would be advantageous over the others in this task.

The tokens are the following:\\
The padding token, the start-of-sequence token, the end-of-sequence token.
\begin{tiny}
\begin{verbatim}
    "<PAD>": 0,
    "<SOS>": 1,
    "<EOS>": 2,
\end{verbatim}
\end{tiny}
The digits of the decimal system (used in both sequence A and B)
\begin{tiny}
\begin{verbatim}
    "0": 3,
    "1": 4,
    "2": 5,
    "3": 6,
    "4": 7,
    "5": 8,
    "6": 9,
    "7": 10,
    "8": 11,
    "9": 12,
\end{verbatim}
\end{tiny}
Operators for defining the quantum state in position (a-h) and mode (x-y)
\begin{tiny}
\begin{verbatim}
    "ax": 13,
    "bx": 14,
    "cx": 15,
    "dx": 16,
    "ex": 17,
    "fx": 18,
    "gx": 19,
    "hx": 20,
    "ay": 21,
    "by": 22,
    "cy": 23,
    "dy": 24,
    "ey": 25,
    "fy": 26,
    "gy": 27,
    "hy": 28,
    "az": 29,
    "bz": 30,
    "cz": 31,
    "dz": 32,
    "ez": 33,
    "fz": 34,
    "gz": 35,
    "hz": 36,
\end{verbatim}
\end{tiny}
Additional symbols for defining the quantum states and the pipe character for separating the three quantum states in sequence A.
\begin{tiny}
\begin{verbatim}
    "[": 37,
    "]": 38,
    "|": 39,
\end{verbatim}
\end{tiny}
Mathematical operators (used both in sequence A and B)
\begin{tiny}
\begin{verbatim}
    "+": 40,
    "-": 41,
    "*": 42,
\end{verbatim}
\end{tiny}
Characters and strings used for python code
\begin{tiny}
\begin{verbatim}
    "for ii in range(N):": 43,
    "\n": 44,
    "    ": 45,
    "e(": 46,
    ")": 47,
    ",": 48,
    "N": 49,
    "ii": 50
\end{verbatim}
\end{tiny}

\begin{figure}[H]
    \centering
    \includegraphics[width = 0.9\textwidth]{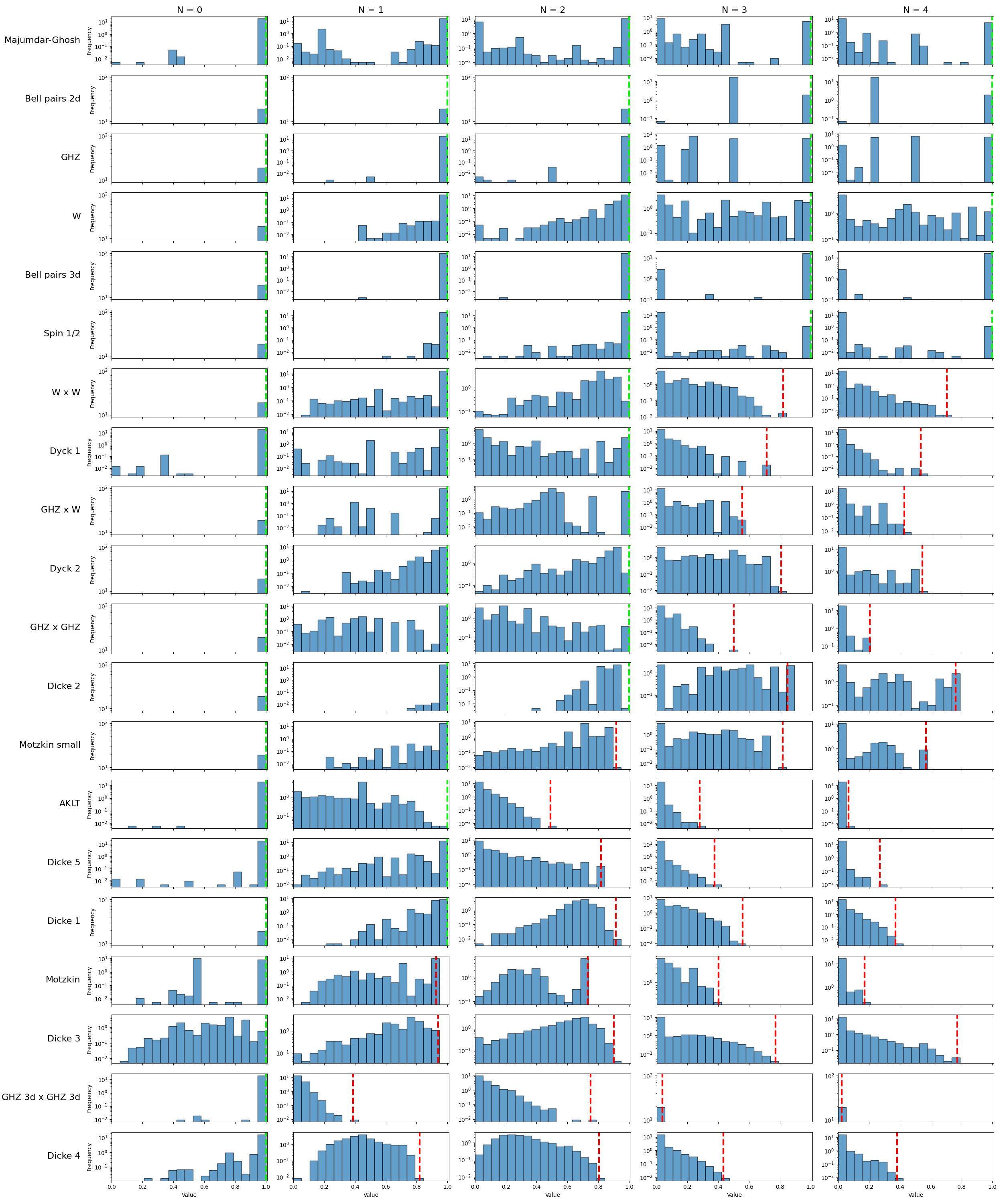}
    \caption{\textbf{Fidelities for codes produced during sampling -} Each row corresponds to a target state class. Each column corresponds to the value of $N$ for which the code is evaluated. Columns $N=3$ and $N=4$ show how the codes perform at generalizing beyond the training range $N=0,1,2$}
    \label{fig:hist}
\end{figure}

\section*{Fidelity computation}
\begin{figure*}
    \centering
    \includegraphics[width = 0.98\textwidth]{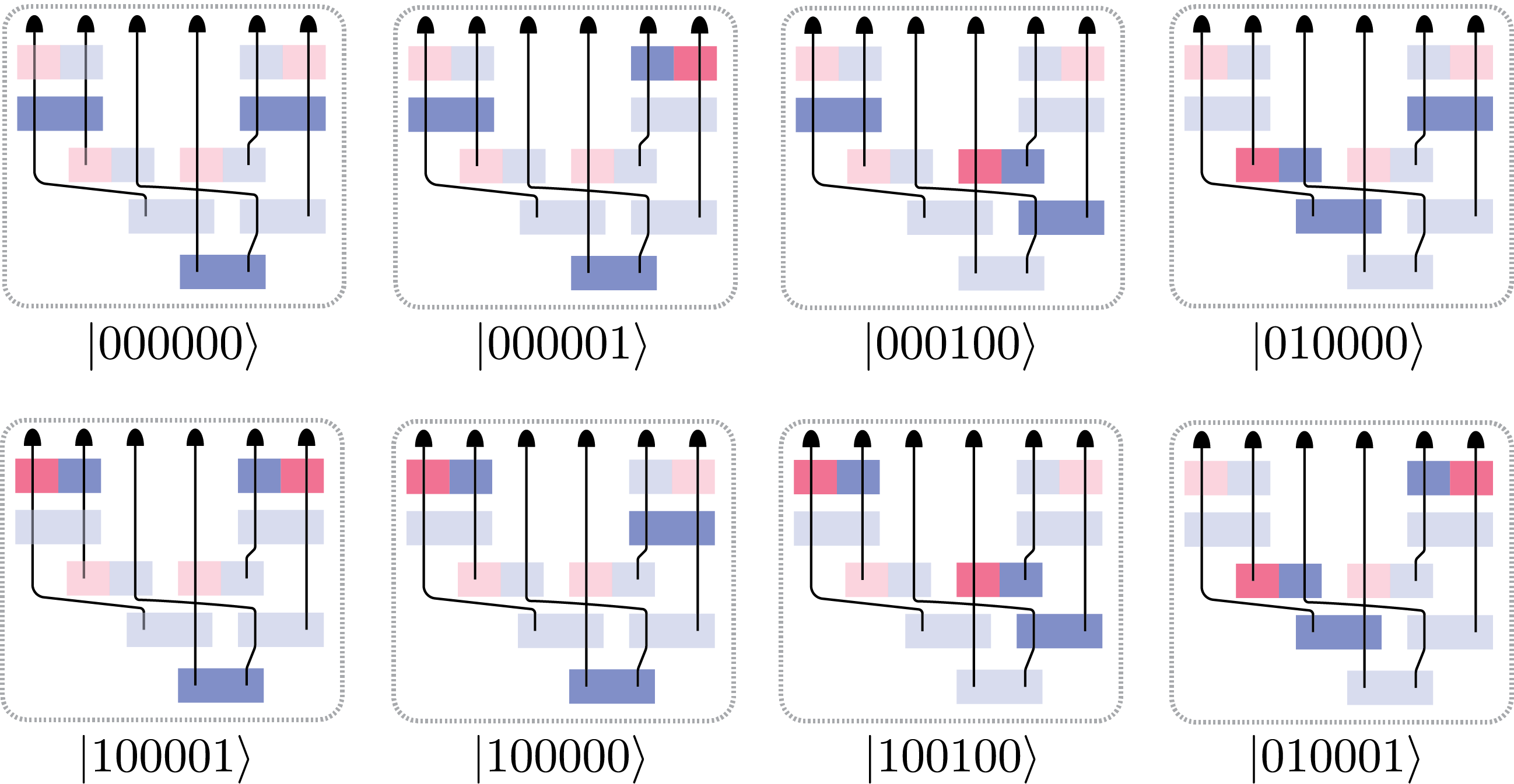}
    \caption{\textbf{Example for computing the resulting quantum state for a given setup.} We show the setup for the spin 1/2 state for six particles. Each panel highlights one configuration of photon-pair sources which can fire at the same time to result in one photon at each detector. All possible configurations are shown. When the six detectors at the top click, we can not know from which configuration of pair-sources the photons have originated. This results in a superposition of all eight possibilities. As different sources can create photons in different modes (blue creating photons in mode 0 and red creating photons in mode 1), the outcome is an entangled state, particularily the six particle spin 1/2 state. The experimental setup is shown with the detectors reordered for visual simplicity. This induces a global permutation of the 0s and 1s of the terms, which can be reversed.}
    \label{fig:fidelitycomp}
\end{figure*}

The fidelity of a quantum state with respect to a target quantum state is a measure of how much the two states overlap. Quantum states are generally represented as elements of a Hilbert space, a vector space equipped with additional properties. The highest possible fidelity is achieved when the two states are exactly the same, giving a value of 1 (the inner product of two identical normalized vectors). The smallest possible value is when the vectors are exactly orthogonal. For example, we define the following one-photon states in bra-ket notation.
\begin{align*}
\ket{\psi_1}&=\ket{0}\\
\ket{\psi_2}&=\ket{1}\\
\ket{\psi_3}&=\frac{1}{\sqrt{2}}(\ket{0}+\ket{1})
\end{align*}
We can compute the fidelity of each state with respect to another state. The fidelities of each state with respect to $\ket{\psi_1}$ are
\begin{align*}
    F_{\ket{\psi_1}}(\ket{\psi_1}) &= |\bra{0}\ket{0}|^2 = 1\\
    F_{\ket{\psi_1}}(\ket{\psi_2}) &= |\bra{0}\ket{1}|^2 = 0\\
    F_{\ket{\psi_1}}(\ket{\psi_3}) &= |\frac{1}{\sqrt{2}}\bra{0}\ket{0}+\frac{1}{\sqrt{2}}\bra{0}\ket{1}|^2 = |\frac{1}{\sqrt{2}}+0|^2 = \frac{1}{2}
\end{align*}
For a multi-photon example, we define the following two normalized four-particle states
\begin{align*}
    \ket{GHZ_4}=\frac{1}{\sqrt{2}}(\ket{0000}+\ket{1111})
    \ket{\psi} = \frac{1}{\sqrt{2}}\ket{0000}+\frac{1}{2}\ket{1111}+\frac{1}{2}\ket{1100}
\end{align*}
The state $\ket{\psi}$ contains the two terms of the $\ket{GHZ_4}$ state, but also another third term, which does not contribute to the inner product performed for computing the fidelity
\begin{align*}
    F_{\ket{GHZ_4}}(\ket{\psi})=|\frac{1}{\sqrt{2}}\frac{1}{\sqrt{2}}\bra{0000}\ket{0000}+\frac{1}{\sqrt{2}}\frac{1}{2}\bra{1111}\ket{1111}+0|^2 = |\frac{1}{2}+\frac{1}{2\sqrt{2}}|^2=\frac{3+2\sqrt{2}}{8}
\end{align*}
In Fig. \ref{fig:fidelitycomp} we show how the terms of a resulting quantum state are derived from a given experimental setup. We compute the the fidelity by overlapping this state with the corresponding target state in the same way as demonstrated here.

\clearpage
\section*{Target classes}
In the following table we show the hand-picked targets for the main task.\\
Some of our target states use ancillary particles. This means that they effectively have a smaller particle size and additional particles that are always in mode $x$ are added in the end to produce a total particle count of $2+2N$. To emphasize the use of ancillary particles, we have colored the parts of the states belonging to ancillary particles in red.

\begin{table}[H]
\begin{tabular}{V{4cm}|V{1cm}|V{10cm}|V{2cm}|V{2cm}}
State Name & size & Quantum State string & correct states & previously known \\
\hline
\hline
Spin 1/2 & 4 &  +1[\textcolor{black}{xxx}\textcolor{red}{x}] +1[\textcolor{black}{xxy}\textcolor{red}{x}] +1[\textcolor{black}{xyx}\textcolor{red}{x}] +1[\textcolor{black}{yxx}\textcolor{red}{x}] +1[\textcolor{black}{yxy}\textcolor{red}{x}]& $\infty$& \textbf{unknown} \\ 
\cline{2-3}
 & 6 &  +1[\textcolor{black}{xxxx}\textcolor{red}{xx}] +1[\textcolor{black}{xxxy}\textcolor{red}{xx}] +1[\textcolor{black}{xxyx}\textcolor{red}{xx}] +1[\textcolor{black}{xyxx}\textcolor{red}{xx}] +1[\textcolor{black}{xyxy}\textcolor{red}{xx}] +1[\textcolor{black}{yxxx}\textcolor{red}{xx}] +1[\textcolor{black}{yxxy}\textcolor{red}{xx}] +1[\textcolor{black}{yxyx}\textcolor{red}{xx}]& &  \\ 
\cline{2-3}
 & 8 &  +1[\textcolor{black}{xxxxx}\textcolor{red}{xxx}] +1[\textcolor{black}{xxxxy}\textcolor{red}{xxx}] +1[\textcolor{black}{xxxyx}\textcolor{red}{xxx}] +1[\textcolor{black}{xxyxx}\textcolor{red}{xxx}] +1[\textcolor{black}{xxyxy}\textcolor{red}{xxx}] +1[\textcolor{black}{xyxxx}\textcolor{red}{xxx}] +1[\textcolor{black}{xyxxy}\textcolor{red}{xxx}] +1[\textcolor{black}{xyxyx}\textcolor{red}{xxx}] +1[\textcolor{black}{yxxxx}\textcolor{red}{xxx}] +1[\textcolor{black}{yxxxy}\textcolor{red}{xxx}] +1[\textcolor{black}{yxxyx}\textcolor{red}{xxx}] +1[\textcolor{black}{yxyxx}\textcolor{red}{xxx}] +1[\textcolor{black}{yxyxy}\textcolor{red}{xxx}]& &  \\ 
\cline{2-3}
\hline
Majumdar-Ghosh & 4 &  -1[xxyy] +2[xyxy] -1[xyyx] -1[yxxy] +2[yxyx] -1[yyxx]& $\infty$& \textbf{unknown} \\ 
\cline{2-3}
 & 6 &  -1[xxyxyy] +1[xxyyxy] +1[xyxxyy] -1[xyxyyx] -1[xyyxxy] +1[xyyxyx] -1[yxxyxy] +1[yxxyyx] +1[yxyxxy] -1[yxyyxx] -1[yyxxyx] +1[yyxyxx]& &  \\ 
\cline{2-3}
 & 8 &  -1[xxyxyxyy] +1[xxyxyyxy] +1[xxyyxxyy] -1[xxyyxyxy] +1[xyxxyxyy] -1[xyxxyyxy] -1[xyxyxxyy] +2[xyxyxyxy] -1[xyxyxyyx] -1[xyxyyxxy] +1[xyxyyxyx] -1[xyyxxyxy] +1[xyyxxyyx] +1[xyyxyxxy] -1[xyyxyxyx] -1[yxxyxyxy] +1[yxxyxyyx] +1[yxxyyxxy] -1[yxxyyxyx] +1[yxyxxyxy] -1[yxyxxyyx] -1[yxyxyxxy] +2[yxyxyxyx] -1[yxyxyyxx] -1[yxyyxxyx] +1[yxyyxyxx] -1[yyxxyxyx] +1[yyxxyyxx] +1[yyxyxxyx] -1[yyxyxyxx]& &  \\ 
\cline{2-3}
\hline
Bell pairs 2d & 4 &  +1[xxxx] +1[xxyy] +1[yyxx] +1[yyyy]& $\infty$& known \\ 
\cline{2-3}
 & 6 &  +1[xxxxxx] +1[xxxxyy] +1[xxyyxx] +1[xxyyyy] +1[yyxxxx] +1[yyxxyy] +1[yyyyxx] +1[yyyyyy]& &  \\ 
\cline{2-3}
 & 8 &  +1[xxxxxxxx] +1[xxxxxxyy] +1[xxxxyyxx] +1[xxxxyyyy] +1[xxyyxxxx] +1[xxyyxxyy] +1[xxyyyyxx] +1[xxyyyyyy] +1[yyxxxxxx] +1[yyxxxxyy] +1[yyxxyyxx] +1[yyxxyyyy] +1[yyyyxxxx] +1[yyyyxxyy] +1[yyyyyyxx] +1[yyyyyyyy]& &  \\ 
\cline{2-3}
\hline
Bell pairs 3d & 4 &  +1[\textcolor{black}{xx}\textcolor{red}{xx}] +1[\textcolor{black}{yy}\textcolor{red}{xx}] +1[\textcolor{black}{zz}\textcolor{red}{xx}]& $\infty$& known \\ 
\cline{2-3}
 & 6 &  +1[\textcolor{black}{xxxx}\textcolor{red}{xx}] +1[\textcolor{black}{xxyy}\textcolor{red}{xx}] +1[\textcolor{black}{xxzz}\textcolor{red}{xx}] +1[\textcolor{black}{yyxx}\textcolor{red}{xx}] +1[\textcolor{black}{yyyy}\textcolor{red}{xx}] +1[\textcolor{black}{yyzz}\textcolor{red}{xx}] +1[\textcolor{black}{zzxx}\textcolor{red}{xx}] +1[\textcolor{black}{zzyy}\textcolor{red}{xx}] +1[\textcolor{black}{zzzz}\textcolor{red}{xx}]& &  \\ 
\cline{2-3}
 & 8 &  +1[\textcolor{black}{xxxxxx}\textcolor{red}{xx}] +1[\textcolor{black}{xxxxyy}\textcolor{red}{xx}] +1[\textcolor{black}{xxxxzz}\textcolor{red}{xx}] +1[\textcolor{black}{xxyyxx}\textcolor{red}{xx}] +1[\textcolor{black}{xxyyyy}\textcolor{red}{xx}] +1[\textcolor{black}{xxyyzz}\textcolor{red}{xx}] +1[\textcolor{black}{xxzzxx}\textcolor{red}{xx}] +1[\textcolor{black}{xxzzyy}\textcolor{red}{xx}] +1[\textcolor{black}{xxzzzz}\textcolor{red}{xx}] +1[\textcolor{black}{yyxxxx}\textcolor{red}{xx}] +1[\textcolor{black}{yyxxyy}\textcolor{red}{xx}] +1[\textcolor{black}{yyxxzz}\textcolor{red}{xx}] +1[\textcolor{black}{yyyyxx}\textcolor{red}{xx}] +1[\textcolor{black}{yyyyyy}\textcolor{red}{xx}] +1[\textcolor{black}{yyyyzz}\textcolor{red}{xx}] +1[\textcolor{black}{yyzzxx}\textcolor{red}{xx}] +1[\textcolor{black}{yyzzyy}\textcolor{red}{xx}] +1[\textcolor{black}{yyzzzz}\textcolor{red}{xx}] +1[\textcolor{black}{zzxxxx}\textcolor{red}{xx}] +1[\textcolor{black}{zzxxyy}\textcolor{red}{xx}] +1[\textcolor{black}{zzxxzz}\textcolor{red}{xx}] +1[\textcolor{black}{zzyyxx}\textcolor{red}{xx}] +1[\textcolor{black}{zzyyyy}\textcolor{red}{xx}] +1[\textcolor{black}{zzyyzz}\textcolor{red}{xx}] +1[\textcolor{black}{zzzzxx}\textcolor{red}{xx}] +1[\textcolor{black}{zzzzyy}\textcolor{red}{xx}] +1[\textcolor{black}{zzzzzz}\textcolor{red}{xx}]& &  \\ 
\cline{2-3}
\hline
GHZ & 4 &  +1[xxxx] +1[yyyy]& $\infty$& known \\ 
\cline{2-3}
 & 6 &  +1[xxxxxx] +1[yyyyyy]& &  \\ 
\cline{2-3}
 & 8 &  +1[xxxxxxxx] +1[yyyyyyyy]& &  \\ 
\cline{2-3}
\hline
W & 4 &  +1[xxxy] +1[xxyx] +1[xyxx] +1[yxxx]& $\infty$& known \\ 
\cline{2-3}
 & 6 &  +1[xxxxxy] +1[xxxxyx] +1[xxxyxx] +1[xxyxxx] +1[xyxxxx] +1[yxxxxx]& &  \\ 
\cline{2-3}
 & 8 &  +1[xxxxxxxy] +1[xxxxxxyx] +1[xxxxxyxx] +1[xxxxyxxx] +1[xxxyxxxx] +1[xxyxxxxx] +1[xyxxxxxx] +1[yxxxxxxx]& &  \\ 
\cline{2-3}
\hline
\end{tabular}
                    
\caption{Classes of quantum states which are of interest in different areas of quantum physics. For each class, the first three states (four, six and eight particles) are shown as strings in the same way that they are used to prompt the model for the main task. The column "correct states" shows, up to which index $N$ the best model output matches the target (the first $N$ states are correct). An infinity sign $\infty$ means, that the meta-solution perfectly matches the target.}
\end{table}
                    
\begin{table}[H]
\begin{tabular}{V{4cm}|V{1cm}|V{10cm}|V{2cm}|V{2cm}}
State Name & size & Quantum State string & correct states & previously known \\
\hline
\hline
GHZ x W & 4 &  +1[xxxy] +1[xxyx] +1[yyxy] +1[yyyx]& 3& unknown \\ 
\cline{2-3}
 & 6 &  +1[xxxxxy] +1[xxxxyx] +1[xxxyxx] +1[yyyxxy] +1[yyyxyx] +1[yyyyxx]& &  \\ 
\cline{2-3}
 & 8 &  +1[xxxxxxxy] +1[xxxxxxyx] +1[xxxxxyxx] +1[xxxxyxxx] +1[yyyyxxxy] +1[yyyyxxyx] +1[yyyyxyxx] +1[yyyyyxxx]& &  \\ 
\cline{2-3}
\hline
W x W & 4 &  +1[xyxy] +1[xyyx] +1[yxxy] +1[yxyx]& 3& unknown \\ 
\cline{2-3}
 & 6 &  +1[xxyxxy] +1[xxyxyx] +1[xxyyxx] +1[xyxxxy] +1[xyxxyx] +1[xyxyxx] +1[yxxxxy] +1[yxxxyx] +1[yxxyxx]& &  \\ 
\cline{2-3}
 & 8 &  +1[xxxyxxxy] +1[xxxyxxyx] +1[xxxyxyxx] +1[xxxyyxxx] +1[xxyxxxxy] +1[xxyxxxyx] +1[xxyxxyxx] +1[xxyxyxxx] +1[xyxxxxxy] +1[xyxxxxyx] +1[xyxxxyxx] +1[xyxxyxxx] +1[yxxxxxxy] +1[yxxxxxyx] +1[yxxxxyxx] +1[yxxxyxxx]& &  \\ 
\cline{2-3}
\hline
Dicke 2 & 4 &  +1[\textcolor{black}{xzz}\textcolor{red}{x}] +1[\textcolor{black}{zxz}\textcolor{red}{x}] +1[\textcolor{black}{zzx}\textcolor{red}{x}]& 3& unknown \\ 
\cline{2-3}
 & 6 &  +1[\textcolor{black}{xxzz}\textcolor{red}{xx}] +1[\textcolor{black}{xzxz}\textcolor{red}{xx}] +1[\textcolor{black}{xzzx}\textcolor{red}{xx}] +1[\textcolor{black}{zxxz}\textcolor{red}{xx}] +1[\textcolor{black}{zxzx}\textcolor{red}{xx}] +1[\textcolor{black}{zzxx}\textcolor{red}{xx}]& &  \\ 
\cline{2-3}
 & 8 &  +1[\textcolor{black}{xxxzz}\textcolor{red}{xxx}] +1[\textcolor{black}{xxzxz}\textcolor{red}{xxx}] +1[\textcolor{black}{xxzzx}\textcolor{red}{xxx}] +1[\textcolor{black}{xzxxz}\textcolor{red}{xxx}] +1[\textcolor{black}{xzxzx}\textcolor{red}{xxx}] +1[\textcolor{black}{xzzxx}\textcolor{red}{xxx}] +1[\textcolor{black}{zxxxz}\textcolor{red}{xxx}] +1[\textcolor{black}{zxxzx}\textcolor{red}{xxx}] +1[\textcolor{black}{zxzxx}\textcolor{red}{xxx}] +1[\textcolor{black}{zzxxx}\textcolor{red}{xxx}]& &  \\ 
\cline{2-3}
\hline
GHZ x GHZ & 4 &  +1[xxxx] +1[xxyy] +1[yyxx] +1[yyyy]& 3& unknown \\ 
\cline{2-3}
 & 6 &  +1[xxxxxx] +1[xxxyyy] +1[yyyxxx] +1[yyyyyy]& &  \\ 
\cline{2-3}
 & 8 &  +1[xxxxxxxx] +1[xxxxyyyy] +1[yyyyxxxx] +1[yyyyyyyy]& &  \\ 
\cline{2-3}
\hline
Dyck 2 & 4 &  +1[yyzz] +1[yzyz]& 3& unknown \\ 
\cline{2-3}
 & 6 &  +1[yyyzzz] +1[yyzyzz] +1[yyzzyz] +1[yzyyzz] +1[yzyzyz]& &  \\ 
\cline{2-3}
 & 8 &  +1[yyyyzzzz] +1[yyyzyzzz] +1[yyyzzyzz] +1[yyyzzzyz] +1[yyzyyzzz] +1[yyzyzyzz] +1[yyzyzzyz] +1[yyzzyyzz] +1[yyzzyzyz] +1[yzyyyzzz] +1[yzyyzyzz] +1[yzyyzzyz] +1[yzyzyyzz] +1[yzyzyzyz]& &  \\ 
\cline{2-3}
\hline
Dyck 1 & 4 &  +1[\textcolor{black}{yz}\textcolor{red}{xx}]& 3& unknown \\ 
\cline{2-3}
 & 6 &  +1[\textcolor{black}{yyzz}\textcolor{red}{xx}] +1[\textcolor{black}{yzyz}\textcolor{red}{xx}]& &  \\ 
\cline{2-3}
 & 8 &  +1[\textcolor{black}{yyyzzz}\textcolor{red}{xx}] +1[\textcolor{black}{yyzyzz}\textcolor{red}{xx}] +1[\textcolor{black}{yyzzyz}\textcolor{red}{xx}] +1[\textcolor{black}{yzyyzz}\textcolor{red}{xx}] +1[\textcolor{black}{yzyzyz}\textcolor{red}{xx}]& &  \\ 
\cline{2-3}
\hline
Dicke 1 & 4 &  +1[\textcolor{black}{xz}\textcolor{red}{xx}] +1[\textcolor{black}{zx}\textcolor{red}{xx}]& 2& unknown \\ 
\cline{2-3}
 & 6 &  +1[\textcolor{black}{xxzz}\textcolor{red}{xx}] +1[\textcolor{black}{xzxz}\textcolor{red}{xx}] +1[\textcolor{black}{xzzx}\textcolor{red}{xx}] +1[\textcolor{black}{zxxz}\textcolor{red}{xx}] +1[\textcolor{black}{zxzx}\textcolor{red}{xx}] +1[\textcolor{black}{zzxx}\textcolor{red}{xx}]& &  \\ 
\cline{2-3}
 & 8 &  +1[\textcolor{black}{xxxzzz}\textcolor{red}{xx}] +1[\textcolor{black}{xxzxzz}\textcolor{red}{xx}] +1[\textcolor{black}{xxzzxz}\textcolor{red}{xx}] +1[\textcolor{black}{xxzzzx}\textcolor{red}{xx}] +1[\textcolor{black}{xzxxzz}\textcolor{red}{xx}] +1[\textcolor{black}{xzxzxz}\textcolor{red}{xx}] +1[\textcolor{black}{xzxzzx}\textcolor{red}{xx}] +1[\textcolor{black}{xzzxxz}\textcolor{red}{xx}] +1[\textcolor{black}{xzzxzx}\textcolor{red}{xx}] +1[\textcolor{black}{xzzzxx}\textcolor{red}{xx}] +1[\textcolor{black}{zxxxzz}\textcolor{red}{xx}] +1[\textcolor{black}{zxxzxz}\textcolor{red}{xx}] +1[\textcolor{black}{zxxzzx}\textcolor{red}{xx}] +1[\textcolor{black}{zxzxxz}\textcolor{red}{xx}] +1[\textcolor{black}{zxzxzx}\textcolor{red}{xx}] +1[\textcolor{black}{zxzzxx}\textcolor{red}{xx}] +1[\textcolor{black}{zzxxxz}\textcolor{red}{xx}] +1[\textcolor{black}{zzxxzx}\textcolor{red}{xx}] +1[\textcolor{black}{zzxzxx}\textcolor{red}{xx}] +1[\textcolor{black}{zzzxxx}\textcolor{red}{xx}]& &  \\ 
\cline{2-3}
\hline
Dicke 5 & 4 &  +1[\textcolor{black}{zzz}\textcolor{red}{x}]& 2& unknown \\ 
\cline{2-3}
 & 6 &  +1[\textcolor{black}{xzzz}\textcolor{red}{xx}] +1[\textcolor{black}{zxzz}\textcolor{red}{xx}] +1[\textcolor{black}{zzxz}\textcolor{red}{xx}] +1[\textcolor{black}{zzzx}\textcolor{red}{xx}]& &  \\ 
\cline{2-3}
 & 8 &  +1[\textcolor{black}{xxzzz}\textcolor{red}{xxx}] +1[\textcolor{black}{xzxzz}\textcolor{red}{xxx}] +1[\textcolor{black}{xzzxz}\textcolor{red}{xxx}] +1[\textcolor{black}{xzzzx}\textcolor{red}{xxx}] +1[\textcolor{black}{zxxzz}\textcolor{red}{xxx}] +1[\textcolor{black}{zxzxz}\textcolor{red}{xxx}] +1[\textcolor{black}{zxzzx}\textcolor{red}{xxx}] +1[\textcolor{black}{zzxxz}\textcolor{red}{xxx}] +1[\textcolor{black}{zzxzx}\textcolor{red}{xxx}] +1[\textcolor{black}{zzzxx}\textcolor{red}{xxx}]& &  \\ 
\cline{2-3}
\hline
AKLT & 4 &  -1[\textcolor{black}{xz}\textcolor{red}{xx}] +1[\textcolor{black}{yy}\textcolor{red}{xx}] -1[\textcolor{black}{zx}\textcolor{red}{xx}]& 2& unknown \\ 
\cline{2-3}
 & 6 &  -1[\textcolor{black}{xyz}\textcolor{red}{xxx}] +1[\textcolor{black}{xzy}\textcolor{red}{xxx}] +1[\textcolor{black}{yxz}\textcolor{red}{xxx}] -1[\textcolor{black}{yzx}\textcolor{red}{xxx}] -1[\textcolor{black}{zxy}\textcolor{red}{xxx}] +1[\textcolor{black}{zyx}\textcolor{red}{xxx}]& &  \\ 
\cline{2-3}
 & 8 &  -1[\textcolor{black}{xyyz}\textcolor{red}{xxxx}] +1[\textcolor{black}{xyzy}\textcolor{red}{xxxx}] +2[\textcolor{black}{xzxz}\textcolor{red}{xxxx}] -1[\textcolor{black}{xzyy}\textcolor{red}{xxxx}] +1[\textcolor{black}{yxyz}\textcolor{red}{xxxx}] -1[\textcolor{black}{yxzy}\textcolor{red}{xxxx}] -1[\textcolor{black}{yyxz}\textcolor{red}{xxxx}] +1[\textcolor{black}{yyyy}\textcolor{red}{xxxx}] -1[\textcolor{black}{yyzx}\textcolor{red}{xxxx}] -1[\textcolor{black}{yzxy}\textcolor{red}{xxxx}] +1[\textcolor{black}{yzyx}\textcolor{red}{xxxx}] -1[\textcolor{black}{zxyy}\textcolor{red}{xxxx}] +2[\textcolor{black}{zxzx}\textcolor{red}{xxxx}] +1[\textcolor{black}{zyxy}\textcolor{red}{xxxx}] -1[\textcolor{black}{zyyx}\textcolor{red}{xxxx}]& &  \\ 
\cline{2-3}
\hline
Motzkin small & 4 &  +1[\textcolor{black}{xy}\textcolor{red}{xx}] +1[\textcolor{black}{zz}\textcolor{red}{xx}]& 2& unknown \\ 
\cline{2-3}
 & 6 &  +1[\textcolor{black}{xyz}\textcolor{red}{xxx}] +1[\textcolor{black}{xzy}\textcolor{red}{xxx}] +1[\textcolor{black}{zxy}\textcolor{red}{xxx}] +1[\textcolor{black}{zzz}\textcolor{red}{xxx}]& &  \\ 
\cline{2-3}
 & 8 &  +1[\textcolor{black}{xxyy}\textcolor{red}{xxxx}] +1[\textcolor{black}{xyxy}\textcolor{red}{xxxx}] +1[\textcolor{black}{xyzz}\textcolor{red}{xxxx}] +1[\textcolor{black}{xzyz}\textcolor{red}{xxxx}] +1[\textcolor{black}{xzzy}\textcolor{red}{xxxx}] +1[\textcolor{black}{zxyz}\textcolor{red}{xxxx}] +1[\textcolor{black}{zxzy}\textcolor{red}{xxxx}] +1[\textcolor{black}{zzxy}\textcolor{red}{xxxx}] +1[\textcolor{black}{zzzz}\textcolor{red}{xxxx}]& &  \\ 
\cline{2-3}
\hline

\end{tabular}
\caption{Classes of quantum states which are of interest in different areas of quantum physics. For each class, the first three states (four, six and eight particles) are shown as strings in the same way that they are used to prompt the model for the main task. The column "correct states" shows, up to which index $N$ the best model output matches the target (the first $N$ states are correct). An infinity sign $\infty$ means, that the meta-solution perfectly matches the target.}
\end{table}
                    
\begin{table}[H]
\begin{tabular}{V{4cm}|V{1cm}|V{10cm}|V{2cm}|V{2cm}}
State Name & size & Quantum State string & correct states & previously known \\
\hline
\hline
Dicke 3 & 4 &  +1[\textcolor{black}{xyz}\textcolor{red}{x}] +1[\textcolor{black}{xzy}\textcolor{red}{x}] +1[\textcolor{black}{yxz}\textcolor{red}{x}] +1[\textcolor{black}{yzx}\textcolor{red}{x}] +1[\textcolor{black}{zxy}\textcolor{red}{x}] +1[\textcolor{black}{zyx}\textcolor{red}{x}]& 1& unknown \\ 
\cline{2-3}
 & 6 &  +1[\textcolor{black}{xxyz}\textcolor{red}{xx}] +1[\textcolor{black}{xxzy}\textcolor{red}{xx}] +1[\textcolor{black}{xyxz}\textcolor{red}{xx}] +1[\textcolor{black}{xyzx}\textcolor{red}{xx}] +1[\textcolor{black}{xzxy}\textcolor{red}{xx}] +1[\textcolor{black}{xzyx}\textcolor{red}{xx}] +1[\textcolor{black}{yxxz}\textcolor{red}{xx}] +1[\textcolor{black}{yxzx}\textcolor{red}{xx}] +1[\textcolor{black}{yzxx}\textcolor{red}{xx}] +1[\textcolor{black}{zxxy}\textcolor{red}{xx}] +1[\textcolor{black}{zxyx}\textcolor{red}{xx}] +1[\textcolor{black}{zyxx}\textcolor{red}{xx}]& &  \\ 
\cline{2-3}
 & 8 &  +1[\textcolor{black}{xxxyz}\textcolor{red}{xxx}] +1[\textcolor{black}{xxxzy}\textcolor{red}{xxx}] +1[\textcolor{black}{xxyxz}\textcolor{red}{xxx}] +1[\textcolor{black}{xxyzx}\textcolor{red}{xxx}] +1[\textcolor{black}{xxzxy}\textcolor{red}{xxx}] +1[\textcolor{black}{xxzyx}\textcolor{red}{xxx}] +1[\textcolor{black}{xyxxz}\textcolor{red}{xxx}] +1[\textcolor{black}{xyxzx}\textcolor{red}{xxx}] +1[\textcolor{black}{xyzxx}\textcolor{red}{xxx}] +1[\textcolor{black}{xzxxy}\textcolor{red}{xxx}] +1[\textcolor{black}{xzxyx}\textcolor{red}{xxx}] +1[\textcolor{black}{xzyxx}\textcolor{red}{xxx}] +1[\textcolor{black}{yxxxz}\textcolor{red}{xxx}] +1[\textcolor{black}{yxxzx}\textcolor{red}{xxx}] +1[\textcolor{black}{yxzxx}\textcolor{red}{xxx}] +1[\textcolor{black}{yzxxx}\textcolor{red}{xxx}] +1[\textcolor{black}{zxxxy}\textcolor{red}{xxx}] +1[\textcolor{black}{zxxyx}\textcolor{red}{xxx}] +1[\textcolor{black}{zxyxx}\textcolor{red}{xxx}] +1[\textcolor{black}{zyxxx}\textcolor{red}{xxx}]& &  \\ 
\cline{2-3}
\hline
Dicke 4 & 4 &  +1[xxyy] +1[xyxy] +1[xyyx] +1[yxxy] +1[yxyx] +1[yyxx]& 1& unknown \\ 
\cline{2-3}
 & 6 &  +1[xxxxyy] +1[xxxyxy] +1[xxxyyx] +1[xxyxxy] +1[xxyxyx] +1[xxyyxx] +1[xyxxxy] +1[xyxxyx] +1[xyxyxx] +1[xyyxxx] +1[yxxxxy] +1[yxxxyx] +1[yxxyxx] +1[yxyxxx] +1[yyxxxx]& &  \\ 
\cline{2-3}
 & 8 &  +1[xxxxxxyy] +1[xxxxxyxy] +1[xxxxxyyx] +1[xxxxyxxy] +1[xxxxyxyx] +1[xxxxyyxx] +1[xxxyxxxy] +1[xxxyxxyx] +1[xxxyxyxx] +1[xxxyyxxx] +1[xxyxxxxy] +1[xxyxxxyx] +1[xxyxxyxx] +1[xxyxyxxx] +1[xxyyxxxx] +1[xyxxxxxy] +1[xyxxxxyx] +1[xyxxxyxx] +1[xyxxyxxx] +1[xyxyxxxx] +1[xyyxxxxx] +1[yxxxxxxy] +1[yxxxxxyx] +1[yxxxxyxx] +1[yxxxyxxx] +1[yxxyxxxx] +1[yxyxxxxx] +1[yyxxxxxx]& &  \\ 
\cline{2-3}
\hline
GHZ 3d x GHZ 3d & 4 &  +1[xxxx] +1[xxyy] +1[xxzz] +1[yyxx] +1[yyyy] +1[yyzz] +1[zzxx] +1[zzyy] +1[zzzz]& 1& unknown \\ 
\cline{2-3}
 & 6 &  +1[xxxxxx] +1[xxxyyy] +1[xxxzzz] +1[yyyxxx] +1[yyyyyy] +1[yyyzzz] +1[zzzxxx] +1[zzzyyy] +1[zzzzzz]& &  \\ 
\cline{2-3}
 & 8 &  +1[xxxxxxxx] +1[xxxxyyyy] +1[xxxxzzzz] +1[yyyyxxxx] +1[yyyyyyyy] +1[yyyyzzzz] +1[zzzzxxxx] +1[zzzzyyyy] +1[zzzzzzzz]& &  \\ 
\cline{2-3}
\hline
Motzkin & 4 &  +1[\textcolor{black}{xyz}\textcolor{red}{x}] +1[\textcolor{black}{xzy}\textcolor{red}{x}] +1[\textcolor{black}{zxy}\textcolor{red}{x}] +1[\textcolor{black}{zzz}\textcolor{red}{x}]& 1& unknown \\ 
\cline{2-3}
 & 6 &  +1[\textcolor{black}{xxyy}\textcolor{red}{xx}] +1[\textcolor{black}{xyxy}\textcolor{red}{xx}] +1[\textcolor{black}{xyzz}\textcolor{red}{xx}] +1[\textcolor{black}{xzyz}\textcolor{red}{xx}] +1[\textcolor{black}{xzzy}\textcolor{red}{xx}] +1[\textcolor{black}{zxyz}\textcolor{red}{xx}] +1[\textcolor{black}{zxzy}\textcolor{red}{xx}] +1[\textcolor{black}{zzxy}\textcolor{red}{xx}] +1[\textcolor{black}{zzzz}\textcolor{red}{xx}]& &  \\ 
\cline{2-3}
 & 8 &  +1[\textcolor{black}{xxyyz}\textcolor{red}{xxx}] +1[\textcolor{black}{xxyzy}\textcolor{red}{xxx}] +1[\textcolor{black}{xxzyy}\textcolor{red}{xxx}] +1[\textcolor{black}{xyxyz}\textcolor{red}{xxx}] +1[\textcolor{black}{xyxzy}\textcolor{red}{xxx}] +1[\textcolor{black}{xyzxy}\textcolor{red}{xxx}] +1[\textcolor{black}{xyzzz}\textcolor{red}{xxx}] +1[\textcolor{black}{xzxyy}\textcolor{red}{xxx}] +1[\textcolor{black}{xzyxy}\textcolor{red}{xxx}] +1[\textcolor{black}{xzyzz}\textcolor{red}{xxx}] +1[\textcolor{black}{xzzyz}\textcolor{red}{xxx}] +1[\textcolor{black}{xzzzy}\textcolor{red}{xxx}] +1[\textcolor{black}{zxxyy}\textcolor{red}{xxx}] +1[\textcolor{black}{zxyxy}\textcolor{red}{xxx}] +1[\textcolor{black}{zxyzz}\textcolor{red}{xxx}] +1[\textcolor{black}{zxzyz}\textcolor{red}{xxx}] +1[\textcolor{black}{zxzzy}\textcolor{red}{xxx}] +1[\textcolor{black}{zzxyz}\textcolor{red}{xxx}] +1[\textcolor{black}{zzxzy}\textcolor{red}{xxx}] +1[\textcolor{black}{zzzxy}\textcolor{red}{xxx}] +1[\textcolor{black}{zzzzz}\textcolor{red}{xxx}]& &  \\ 
\cline{2-3}
\hline
\end{tabular}
\caption{Classes of quantum states which are of interest in different areas of quantum physics. For each class, the first three states (four, six and eight particles) are shown as strings in the same way that they are used to prompt the model for the main task. The column "correct states" shows, up to which index $N$ the best model output matches the target (the first $N$ states are correct). An infinity sign $\infty$ means, that the meta-solution perfectly matches the target.}
\end{table}

\pagebreak

\section{Overlap between target states and training data}
\begin{figure}[H]
    \centering
    \includegraphics[width = 0.8\textwidth]{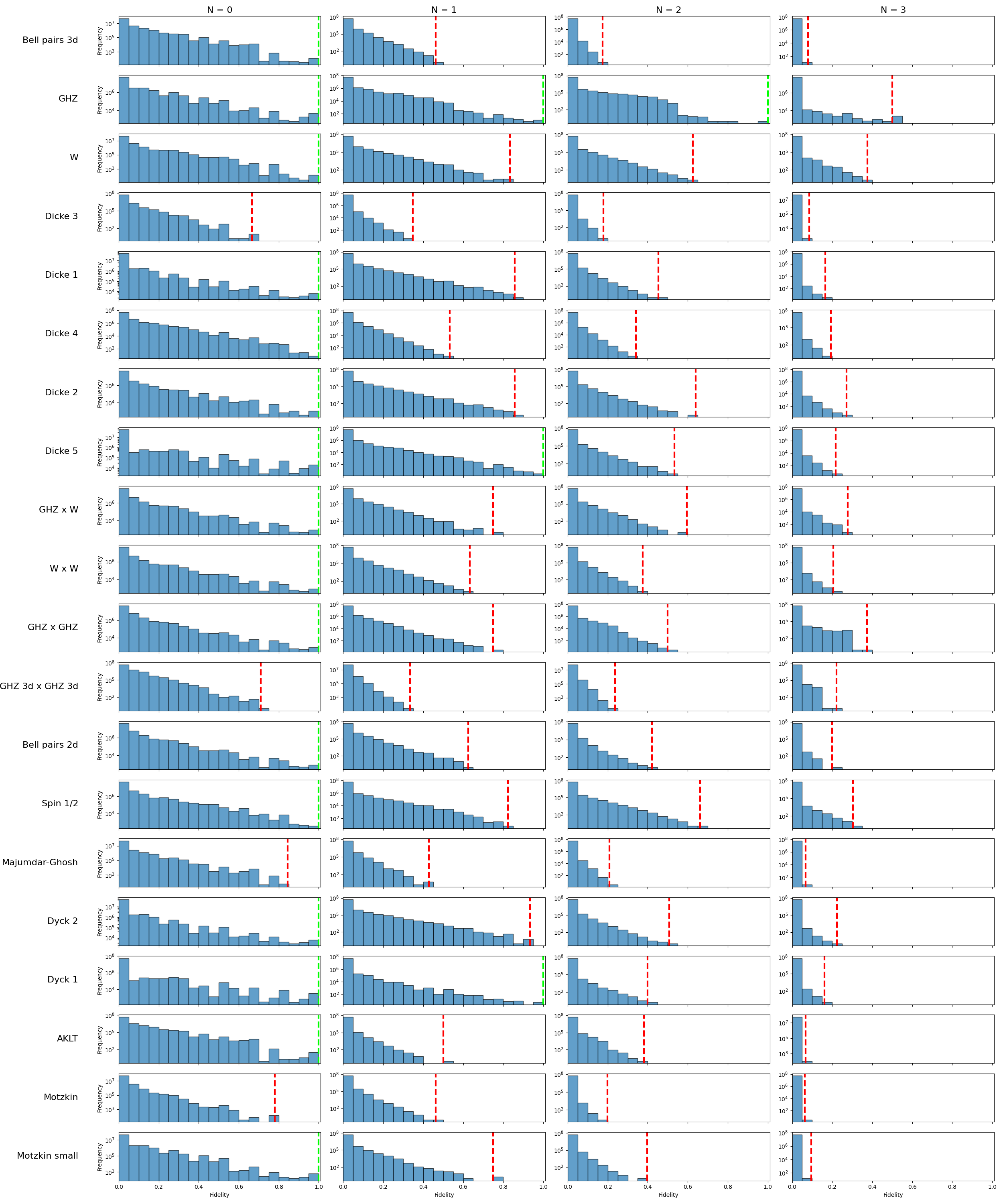}
    \caption{\textbf{Overlap between target states and training data -} We compute the overlap (fidelity) between all target states and the states created by the 56M codes in the training data. For each sample of the training data, sequence A only includes the four-, six- and eight-particle state ($N=0,1,2$), but we can find the ten-particle state corresponding to the code by setting $N=3$ and computing the resulting setup. The highest of all 56M fidelities is emphasized by a vertical dashed line in each histogram (colored green for perfect overlap).}
    \label{fig:hist_rand}
\end{figure}

To evaluate the generalization capabilities of the model, we computed the fidelity of all samples in the training data with respect to all target data and show their distribution in Fig. \ref{fig:hist_rand}. This serves to rule out that the full space of possible solutions is explored. Most of the target states for four particles ($N=0$) are found by codes in the training data. This is not very surprising, as the space of four-particle states is small enough to be fully explored by a large number of samples. Notably, there is even a single training sample code, which generates setups creating the GHZ states for $N=0,1,2$, but it does not match the GHZ state for $N=3$ (for which the fidelity is $0.25$). This means that we would not consider this code a complete solution, as it does not generalize to $N\geq 3$.

\begin{figure}[H]
    \centering
    \includegraphics[width = 0.85\textwidth]{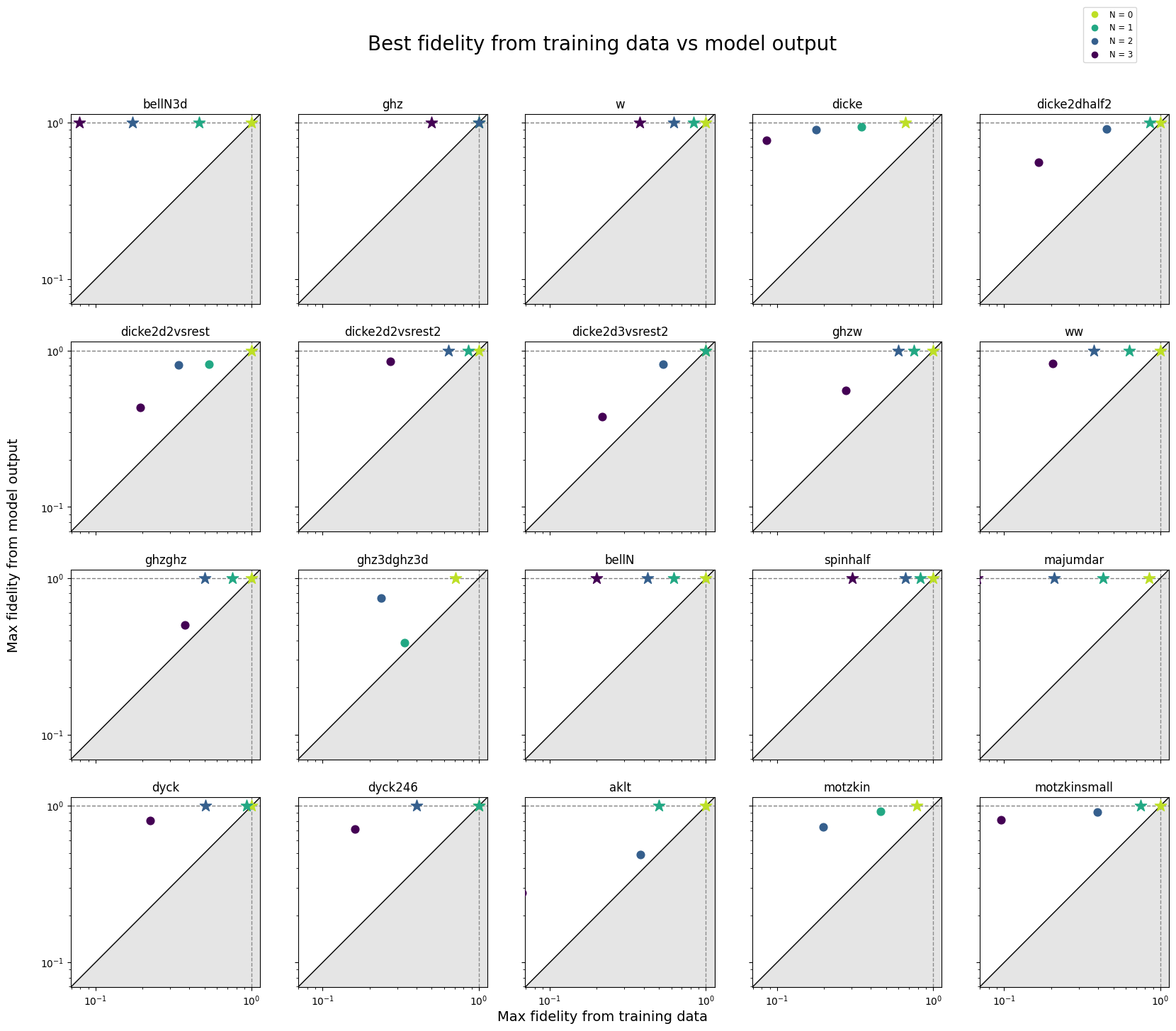}
    \caption{\textbf{Comparing the maximum overlap with the overlap of the best model output -} We compare the best fidelities of codes found in the training data and codes generated by the model for each target class and for $N=0,1,2,3$. Instances where the model prediction matches the target state exactly are marked with a star. For all points, the best model output is better. We also see that the best values from training codes decrease as $N$ grows, but the model output either decreases slower or even stays constant at fidelity 1 for the six classes where the generalizing code was discovered. The grey coloring under the diagonal shows the area where the best training sample would be better than the best model output.}
    \label{fig:rand_vs_pred}
\end{figure}

\begin{figure}[H]
    \centering
    \includegraphics[width = 0.85\textwidth]{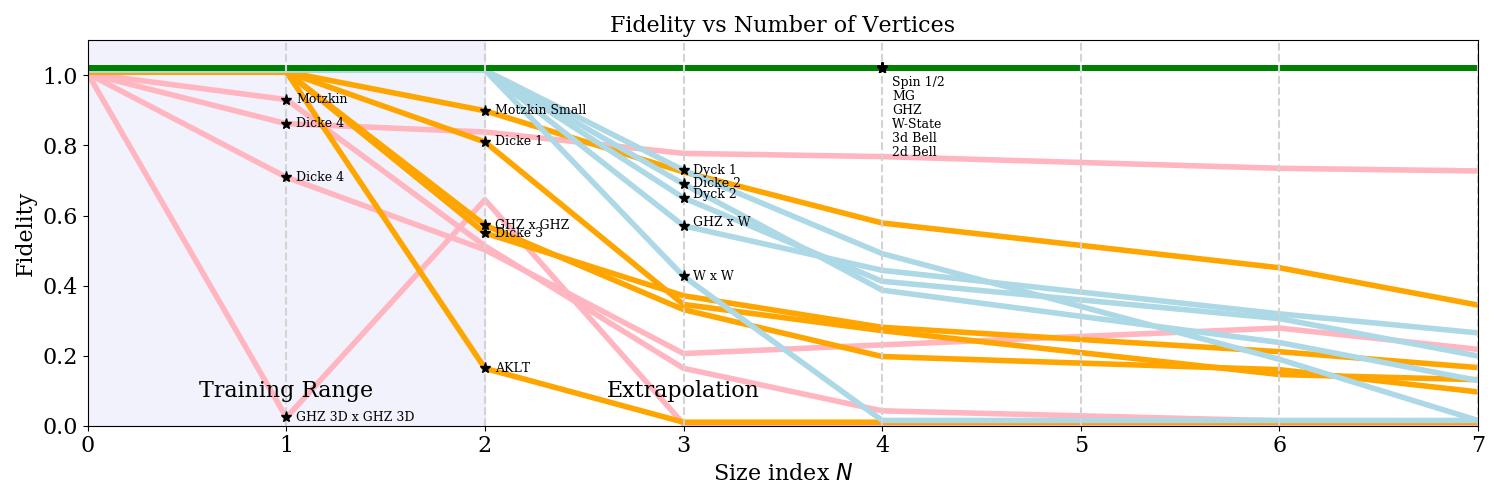}
    \caption{\textbf{Extended version of fidelity vs. particle number -} We show the resulting fidelities of the best produced code for all 20 target classes as an extension of Fig. \ref{fig:fidvsvert} of the main text. Here we include the fidelities up to $N=7$ (18 particles). This shows that the produced codes for the six successful classes (shown in green) are not only correct for small particle numbers but extend beyond.}
    \label{fig:fidsvsvert_extended}
\end{figure}

\pagebreak
\section{Computation of computational cost for standard design algorithm}

To estimate the computational cost of the direct optimization task we choose to calculate the number of floating point operations (FLOP) necessary to compute the loss function. This number is a useful measure as it can be converted to a monetary cost by multiplying it with a factor corresponding to specific hardware.

The loss function for an experiment with $2N$ particles is \cite{pytheus}
\begin{align}
    L(\omega) = |\braket{\psi_{\text{target}}|\psi(\omega)}|^2
\end{align}
We can estimate a lower bound for the number of floating point operations by the number of operations necessary to compute just the state $\ket{\psi(\omega)}$ from the parameters $\omega$.
\begin{align}
    FLOP(L(\omega))\geq FLOP(\ket{\psi(\omega)})
\end{align}
$\omega$ is a vector with each entry corresponding to a weight of an edge in a complete bi-colored graph \cite{pytheus}. A quantum state of $N$ particles with dimensionality $k$ can have $k^N$ kets (for $N=4$, $k=2$: $\ket{0000}$, $\ket{0001}$, $\ket{0010}$, ...,  $\ket{1111}$). The computation of the amplitude for each ket requires us to multiply the weights of the edges in a perfect matching for each perfect matching (also referred to as a Hafnian), e.g.
\begin{align}
    c_{0000} = \omega^{00}_{01}*\omega^{00}_{23}+\omega^{00}_{02}*\omega^{00}_{13}+\omega^{00}_{03}*\omega^{00}_{12}
\end{align}
The number of multiplications and additions for each amplitude for a graph with $N$ nodes can be estimated by $(N/2-1)*\#\text{PerfectMatchings}=(N/2)*(N-1)!! -1$ (double factorial).\\
We can thus estimate the number of floating point operations necessary for computing all $k^N$ amplitudes as
\begin{align}
    \text{FLOP}(\ket{\psi(\omega)})\approx (k^N)*(N/2)*(N-1)!!
\end{align}
At this point we want to note that the minimal number of floating point operations necessary for computing the state is lower than the estimation we give here, because some multiplications could be stored and reused. However, the factor converges to a constant value as the number of nodes grows, so the overall scaling is correct. We also note that the computation of a Hafnian is \#P-hard, so no polynomial-size calculation can exist. Our own computation of the quantum state does not reuse any multiplications, so it matches our FLOP calculation exactly.\\
We now can compute the lower bound for computing the loss function for $k=3$ and different values of $N$:
\begin{align*}
N =&~ 4 &\rightarrow&~ 486    \\
N =&~ 6 &\rightarrow&~ 32805 \\
N =&~ 8 &\rightarrow&~ 2755620 \\
N =&~ 20 &\rightarrow&~ 2*10^{19}
\end{align*}

Further, this calculation only considers the computational cost for computing the loss function once. As the system grows, the number of free parameters in the optimization grows with $N_\text{param}=k^2*\frac{N(N-1)}{2}$, so we can assume that we will need to compute the loss function more often because more steps are necessary to find the minimum.

\section{Fine-tuning of a pretrained LLM on the code-generation task}
Large language models (LLMs) are pretrained on a very large corpus of text, including scripts written in a variety of different programming languages. They have been shown to be very capable at complex programming tasks. Furthermore, fine-tuning LLMs - training a base model on a specialized tasks - has been shown to be  successfull in acquiring domain-specific capabilities with small computational cost relatively to pretraining. Therefore it is reasonable to expect that it is possible to fine-tune an existing LLM on our code-generation task. In this section we show that this is possible with a comparable compute budget as training a smaller model from scratch.
\textbf{Model choice}
We chose the 8B parameter version of Llama-3. Models of this size are considered 'small', compared to e.g. the 70B parameter version of Llama-3.

\textbf{Tokenization and Data input}
Llama-3 8B like the majority of LLMs is a decoder-only model (in contrast to our encoder-decoder model) and comes with its own trained tokenizer. Our data thus consists of 56M raw text samples of the form\\
\texttt{<|begin\_of\_text|>+<|start\_header\_id|>[quantum state]<|end\_header\_id|>+state\_str+}\\
\texttt{<|start\_header\_id|>[code]<|end\_header\_id|>+code\_str+<|end\_of\_text|>}.\\
This text is tokenized using the Llama-3 tokenizer and multiple samples are packed together to fill as much of the 8192 token context length of Llama-3 8B.

\textbf{Memory requirements for fine-tuning \textit{Meta-Llama-3-8B} --} This note details the GPU‐memory footprint of \textit{Meta-Llama-3-8B} under two training regimes: (i) conventional full-precision fine-tuning and (ii) parameter-efficient fine-tuning with QLoRA\cite{dettmers2023}.  We assume a single 40 GB GPU and sequence-packing with gradient checkpointing; activations therefore scale linearly with the packed-sequence batch size~$N$ (each sample contains $\approx 8192$ tokens).

\textit{Full-precision fine-tuning --} In its original release the model’s weights are stored in \textsc{bf}16, consuming 15.1 GB (Table~\ref{tab:full}).  During standard fine-tuning the same amount of memory is required for gradients, while the Adam optimiser adds first- and second-moment estimates in 32-bit precision.  Before accounting for activations, a single training step already demands 90.7 GB—more than twice the capacity of the target GPU—necessitating model parallelism or offloading strategies that incur additional overhead.

\begin{table}[h]
\centering
\caption{Memory footprint of a single training step in full-precision.}
\label{tab:full}
\begin{tabular}{@{}lcc@{}}
\toprule
\textbf{Component} & \textbf{Bytes per param} & \textbf{Memory}\\
\midrule
Weights (\textsc{bf}16)                     & 2 & 15.1 GB\\
Gradients (\textsc{bf}16)                   & 2 & 15.1 GB\\
Adam 1\textsuperscript{st} \& 2\textsuperscript{nd} moments (FP32) & 8 & 60.5 GB\\
\addlinespace
\textbf{Static total (excl.\ activations)}  & -- & \textbf{90.7 GB}\\
Activations & -- & $1.5\;{\rm GB}\times N$\\
\bottomrule
\end{tabular}
\end{table}

\textit{QLoRA fine-tuning --} QLoRA combines low-rank adaptation (LoRA) with 4-bit NormalFloat-4 (NF4) quantisation of the frozen backbone.  Only a small fraction of the parameters—the low-rank adapters—remain in \textsc{bf}16 and are updated during training.  The backbone is stored once in 4-bit form, and the optimiser keeps 8-bit PagedAdam states for the trainable parameters.  As summarised in Table~\ref{tab:qlora}, the static memory requirement drops to 5.5 GB, placing the entire training step comfortably within the 40 GB budget even for moderate batch sizes.

\begin{table}[h]
\centering
\caption{Memory footprint of a single training step with QLoRA ($r=64$, 2.5 \% trainable parameters).}
\label{tab:qlora}
\begin{tabular}{@{}lcc@{}}
\toprule
\textbf{Component} & \textbf{Bytes per param} & \textbf{Memory}\\
\midrule
Frozen backbone (4-bit NF4 $\,+$ overhead) & $\approx 0.5$ & 4.3 GB\\
LoRA weights (\textsc{bf}16)               & 2 & 0.41 GB\\
LoRA gradients (\textsc{bf}16)             & 2 & 0.41 GB\\
PagedAdam states (8-bit)                   & 2 & 0.41 GB\\
\addlinespace
\textbf{Static total (excl.\ activations)} & -- & \textbf{5.5 GB}\\
Activations & -- & $1.5\;{\rm GB}\times N$\\
\bottomrule
\end{tabular}
\end{table}

\textit{Practical implication --} With QLoRA, \textit{Meta-Llama-3-8B} can be fine-tuned on a single 40 GB GPU without model sharding or parameter offloading, enabling substantially faster iteration while preserving model quality.

\begin{figure}[H]
    \centering
    \includegraphics[width = 0.85\textwidth]{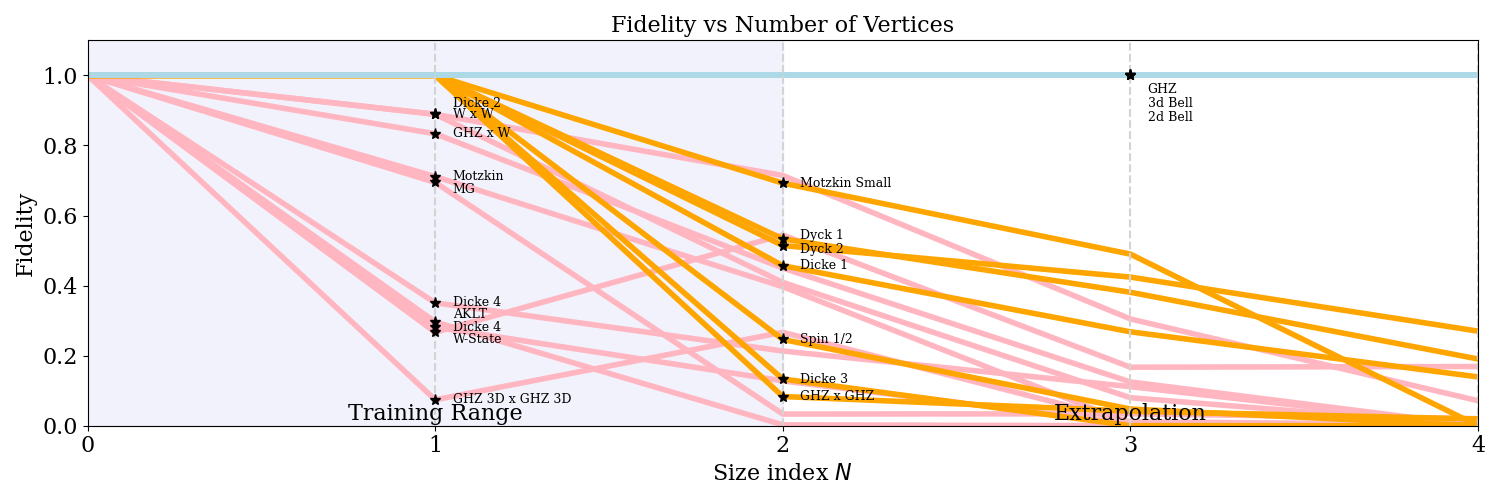}
    \caption{\textbf{Results for fine-tuned \textit{Meta-Llama-3-8B}} We show the resulting fidelities of the best produced code for the 20 target classes in the main task.}
    \label{fig:llama}
\end{figure}

\textbf{Results -- } 
We see in Fig. \ref{fig:llama} that the approach of fine-tuning a larger pretrained model can achieve positive results, but does not outperform our smaller model trained from scratch with the same computational resources. We do see this as a possible avenue for applications where not as much training data is available.

\section{Background on Quantum Circuits}
A quantum circuit is the quantum counterpart to a circuit of logical gates in classical computer science. It can be represented as a diagram that can be read from left to right (see right side of Fig. \ref{fig:cnot}). A quantum circuit is usually initialized with $N$ qubits in a separable state such as $\ket{0}^{\bigotimes N}$. Gates such as \texttt{H} (Hadamard gate) and \texttt{CNOT} are then consequently applied to produce an entangled state. Designing a quantum circuit for a desired target state can be a difficult task and is approached with different algorithmic methods such as reinforcement learning \cite{ostaszewski2021reinforcement, https://doi.org/10.48550/arxiv.2311.18588,https://doi.org/10.48550/arxiv.2402.17761} or variational optimization \cite{Kottmann2023, maclellan2024end}.

\begin{figure}
    \centering
    \includegraphics[width = 0.85\textwidth]{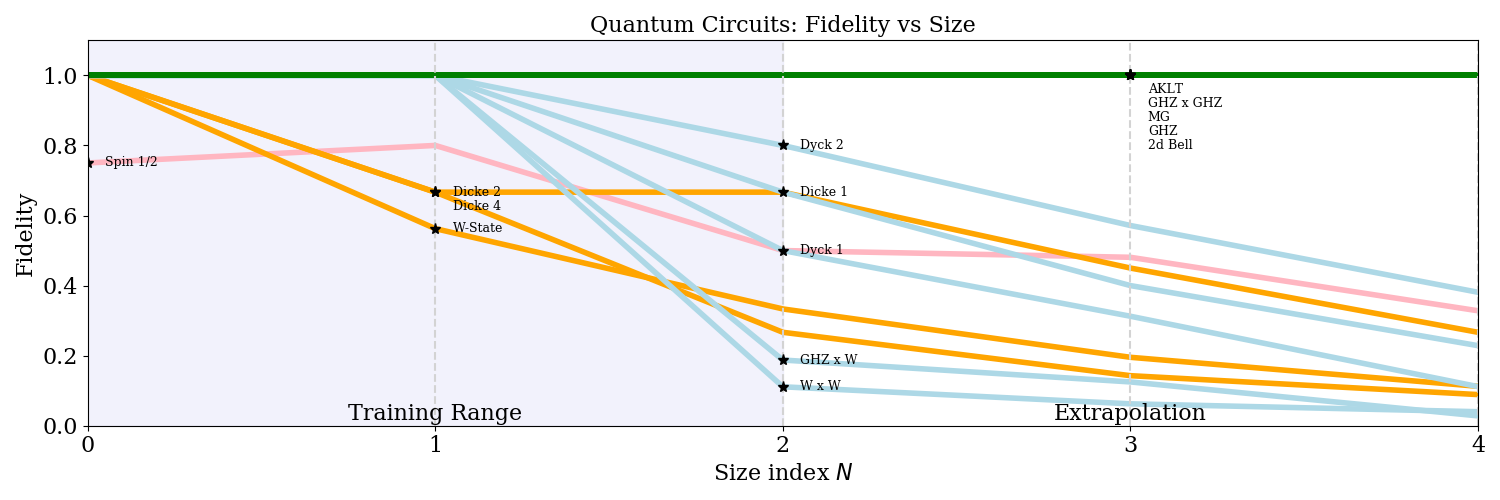}
    \caption{\textbf{Results for quantum circuits -} We show the best fidelities for each of the target classes that apply to the quantum circuit case (only qubits are permitted, so we do not test three-dimensional states).}
    \label{fig:cnot}
\end{figure}

\section{Data generation for additional tasks (quantum circuits and quantum graph states)}
\textbf{Code structure}\\
Each training sample code has the structure:\\
\begin{verbatim}
[1-5 lines of gate functions]
for ii in range([range formula]):
    [1-5 lines of gate functions]
[1-5 lines of gate functions]
\end{verbatim}
The number of lines is picked randomly (uniform distribution). In the case of quantum circuits the type of gate is also picked randomly for each line.\\

\textbf{Valid argument formulas}\\
\texttt{ints = ['-4','-3','-2','-1','','1','2','3','4']}\\
\texttt{scale = ['','+NN','+2*NN']}\\
\texttt{its = ['-ii','','+ii','+2*ii']}
Valid argument formulas for lines outside of the for-loop are concatenated combinations of \texttt{ints} and \texttt{scale} that satisfy $0\leq f(N) \leq 2+2N$ for $N<8$.\\

Valid arguments formulas for lines inside the for-loop are concatenated combinations of \texttt{ints}, \texttt{scale}, and \texttt{its} that satisfy $0\leq f(N) \leq 2+2N$ for $N<8$ and $0\leq ii<\texttt{range}$.\\

For two-qubit gates, we check if the first and second argument of the gate are equal for any valid combination of $N$ and $ii$. If they are, we pick a different argument formula for the second argument, to avoid the invalid operation of applying a two-qubit gate to a single qubit.\\

\textbf{Computation and expression of states}\\
We compute the resulting states by setting the values for $N=0,1,2$ and executing the code with IBM's Qiskit \cite{javadi2024quantum}. We further process the states by filtering out terms with zero amplitudes and dividing all coefficients by the smallest coefficient. This resulted in all states in our training data having integer valued coefficients. To reduce ambugity in the expression of the states, we multiply each state with a global phase such that the first coefficient has a positive sign. We then concatenate the states for $N=0,1,2$ in the same way as it is done in the main task, separated by the token \texttt{<SEP>}.

\newpage
\textbf{Tokenization}\\
We use the same vocabularies for both tasks (circuits and graph states).\\
\begin{tabular}{p{0.48\textwidth} p{0.48\textwidth}}
\textbf{Source Vocabulary (states)} & \textbf{Target Vocabulary (codes)} \\

\begin{verbatim}
{
"<PAD>": 0,
"<SOS>": 1,
"<EOS>": 2,
"0": 3,
"1": 4,
"2": 5,
"3": 6,
"4": 7,
"5": 8,
"6": 9,
"7": 10,
"8": 11,
"9": 12,
"X": 13,
"Y": 14,
"|": 15,
">": 16,
"+": 17,
"-": 18,
"sqrt(": 19,
"/": 20,
"*": 21,
"(": 22,
")": 23,
"<SEP>": 24
}
\end{verbatim}
&
\begin{verbatim}
{
"<PAD>": 0,
"<SOS>": 1,
"<EOS>": 2,
"0": 3,
"1": 4,
"2": 5,
"3": 6,
"4": 7,
"5": 8,
"6": 9,
"7": 10,
"8": 11,
"9": 12,
"for ii in range(": 13,
"):" : 14,
"\n": 15,
"    ": 16,
"qH": 17,
"qCNOT": 18,
"qX": 19,
"qZ": 20,
"+": 21,
"-": 22,
"*": 23,
"NN": 24,
"ii": 25,
"(": 26,
")": 27,
",": 28,
"qToffoli": 29,
"qCSWAP": 30,
"qCZ": 31
}
\end{verbatim}
\end{tabular}

For the quantum circuit design task, we allow a maximum token sequence length of 640 for source and target sequence each. For the quantum graph state design task, the maximum token sequence length for the source sequence (states) is set to 1792, because graph states (including the linear, ring and star graph) commonly contain all possible kets.

\begin{figure*}
    \centering
    \includegraphics[width = 0.7\textwidth]{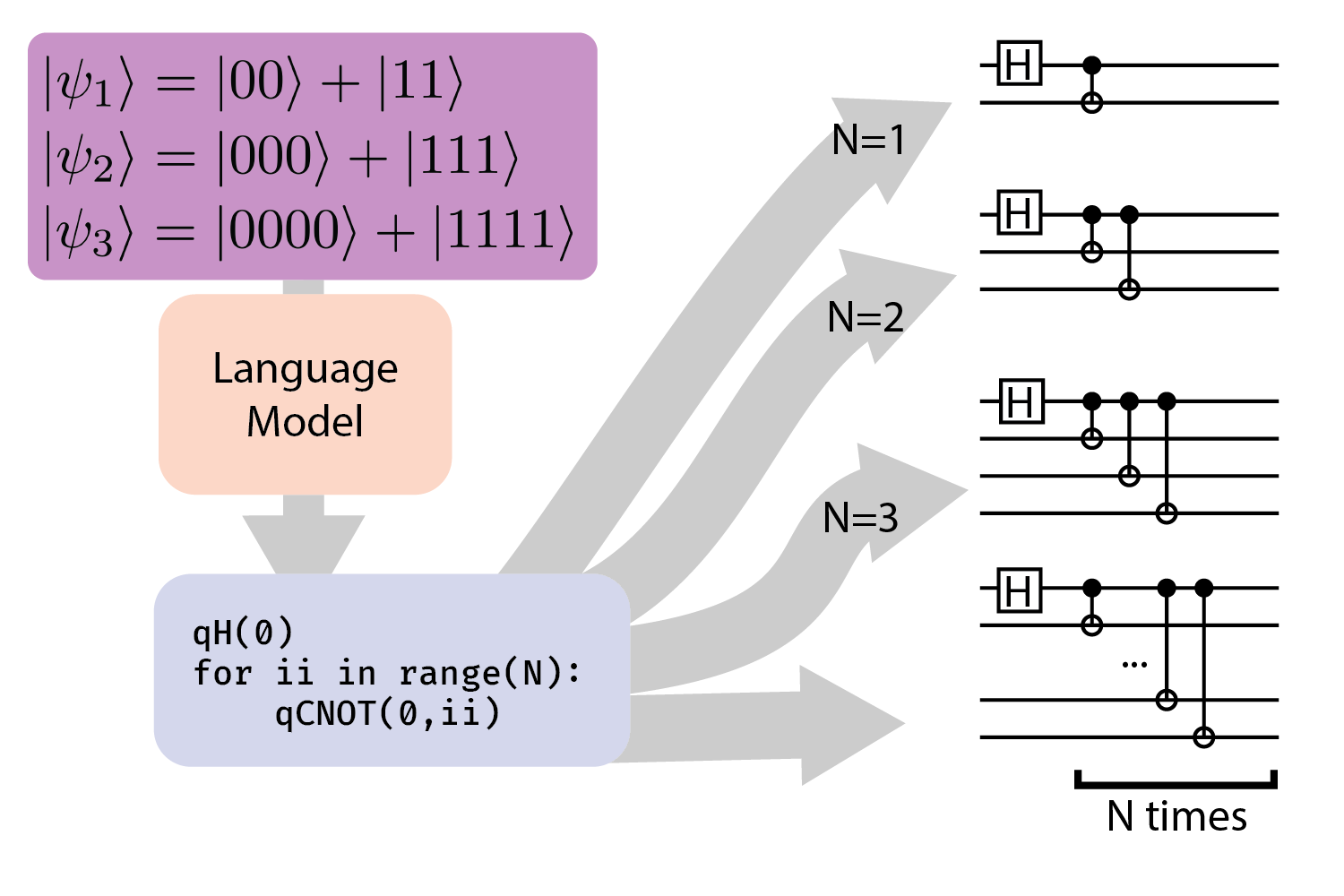}
    \caption{\textbf{Meta-design process for the quantum circuit task.} The trained model receives the GHZ states for two, three, and four particles as input. The output of the model is a code which produces the quantum circuits on the right for varying particle number $N$.}
    \label{fig:cnot}
\end{figure*}
\clearpage
\pagebreak

\begin{table}[H]
\begin{tabular}{V{4cm}|V{1cm}|V{8cm}|V{2cm}|V{2cm}}
State Name & size & Quantum State string & correct states & previously known \\
\hline
\hline
GHZ & 2 & 1$\vert$XX$>$ +1$\vert$YY$>$ & $\infty$& known \\ 
\cline{2-3}
 & 4 & 1$\vert$XXXX$>$ +1$\vert$YYYY$>$ & &  \\ 
\cline{2-3}
 & 6 & 1$\vert$XXXXXX$>$ +1$\vert$YYYYYY$>$ & &  \\ 
\cline{2-3}
\hline
W & 2 & 1$\vert$XY$>$ +1$\vert$YX$>$ & 1& known \\ 
\cline{2-3}
 & 4 & 1$\vert$XXXY$>$ +1$\vert$XXYX$>$ +1$\vert$XYXX$>$ +1$\vert$YXXX$>$ & &  \\ 
\cline{2-3}
 & 6 & 1$\vert$XXXXXY$>$ +1$\vert$XXXXYX$>$ +1$\vert$XXXYXX$>$ +1$\vert$XXYXXX$>$ +1$\vert$XYXXXX$>$ +1$\vert$YXXXXX$>$ & &  \\ 
\cline{2-3}
\hline
Dicke 1 & 2 & 1$\vert$XX$>$ & 2& unknown \\ 
\cline{2-3}
 & 4 & 1$\vert$XYXX$>$ +1$\vert$YXXX$>$ & &  \\ 
\cline{2-3}
 & 6 & 1$\vert$XXYYXX$>$ +1$\vert$XYXYXX$>$ +1$\vert$XYYXXX$>$ +1$\vert$YXXYXX$>$ +1$\vert$YXYXXX$>$ +1$\vert$YYXXXX$>$ & &  \\ 
\cline{2-3}
\hline
Dicke 4 & 2 & 1$\vert$YY$>$ & 1& unknown \\ 
\cline{2-3}
 & 4 & 1$\vert$XXYY$>$ +1$\vert$XYXY$>$ +1$\vert$XYYX$>$ +1$\vert$YXXY$>$ +1$\vert$YXYX$>$ +1$\vert$YYXX$>$ & &  \\ 
\cline{2-3}
 & 6 & 1$\vert$XXXXYY$>$ +1$\vert$XXXYXY$>$ +1$\vert$XXXYYX$>$ +1$\vert$XXYXXY$>$ +1$\vert$XXYXYX$>$ +1$\vert$XXYYXX$>$ +1$\vert$XYXXXY$>$ +1$\vert$XYXXYX$>$ +1$\vert$XYXYXX$>$ +1$\vert$XYYXXX$>$ +1$\vert$YXXXXY$>$ +1$\vert$YXXXYX$>$ +1$\vert$YXXYXX$>$ +1$\vert$YXYXXX$>$ +1$\vert$YYXXXX$>$ & &  \\ 
\cline{2-3}
\hline
Dicke 2 & 2 & 1$\vert$YY$>$ & 1& unknown \\ 
\cline{2-3}
 & 4 & 1$\vert$XYYX$>$ +1$\vert$YXYX$>$ +1$\vert$YYXX$>$ & &  \\ 
\cline{2-3}
 & 6 & 1$\vert$XXYYXX$>$ +1$\vert$XYXYXX$>$ +1$\vert$XYYXXX$>$ +1$\vert$YXXYXX$>$ +1$\vert$YXYXXX$>$ +1$\vert$YYXXXX$>$ & &  \\ 
\cline{2-3}
\hline
GHZ x W & 2 & 1$\vert$XY$>$ +1$\vert$YY$>$ & 2& unknown \\ 
\cline{2-3}
 & 4 & 1$\vert$XXXY$>$ +1$\vert$XXYX$>$ +1$\vert$YYXY$>$ +1$\vert$YYYX$>$ & &  \\ 
\cline{2-3}
 & 6 & 1$\vert$XXXXXY$>$ +1$\vert$XXXXYX$>$ +1$\vert$XXXYXX$>$ +1$\vert$YYYXXY$>$ +1$\vert$YYYXYX$>$ +1$\vert$YYYYXX$>$ & &  \\ 
\cline{2-3}
\hline
\end{tabular}
\caption{\textbf{Target states for quantum circuit design (Part 1)}}
\end{table}
                    
\begin{table}[H]
\begin{tabular}{V{4cm}|V{1cm}|V{8cm}|V{2cm}|V{2cm}}
State Name & size & Quantum State string & correct states & previously known \\
\hline
\hline
W x W & 2 & 1$\vert$YY$>$ & 2& unknown \\ 
\cline{2-3}
 & 4 & 1$\vert$XYXY$>$ +1$\vert$XYYX$>$ +1$\vert$YXXY$>$ +1$\vert$YXYX$>$ & &  \\ 
\cline{2-3}
 & 6 & 1$\vert$XXYXXY$>$ +1$\vert$XXYXYX$>$ +1$\vert$XXYYXX$>$ +1$\vert$XYXXXY$>$ +1$\vert$XYXXYX$>$ +1$\vert$XYXYXX$>$ +1$\vert$YXXXXY$>$ +1$\vert$YXXXYX$>$ +1$\vert$YXXYXX$>$ & &  \\ 
\cline{2-3}
\hline
GHZ x GHZ & 2 & 1$\vert$XX$>$ +1$\vert$XY$>$ +1$\vert$YX$>$ +1$\vert$YY$>$ & $\infty$& known \\ 
\cline{2-3}
 & 4 & 1$\vert$XXXX$>$ +1$\vert$XXYY$>$ +1$\vert$YYXX$>$ +1$\vert$YYYY$>$ & &  \\ 
\cline{2-3}
 & 6 & 1$\vert$XXXXXX$>$ +1$\vert$XXXYYY$>$ +1$\vert$YYYXXX$>$ +1$\vert$YYYYYY$>$ & &  \\ 
\cline{2-3}
\hline
Bell pairs 2d & 2 & 1$\vert$XX$>$ +1$\vert$YY$>$ & $\infty$& known \\ 
\cline{2-3}
 & 4 & 1$\vert$XXXX$>$ +1$\vert$XXYY$>$ +1$\vert$YYXX$>$ +1$\vert$YYYY$>$ & &  \\ 
\cline{2-3}
 & 6 & 1$\vert$XXXXXX$>$ +1$\vert$XXXXYY$>$ +1$\vert$XXYYXX$>$ +1$\vert$XXYYYY$>$ +1$\vert$YYXXXX$>$ +1$\vert$YYXXYY$>$ +1$\vert$YYYYXX$>$ +1$\vert$YYYYYY$>$ & &  \\ 
\cline{2-3}
\hline
Spin 1/2 & 2 & 1$\vert$XX$>$ +1$\vert$XY$>$ +1$\vert$YX$>$ & 0& unknown \\ 
\cline{2-3}
 & 4 & 1$\vert$XXXX$>$ +1$\vert$XXYX$>$ +1$\vert$XYXX$>$ +1$\vert$YXXX$>$ +1$\vert$YXYX$>$ & &  \\ 
\cline{2-3}
 & 6 & 1$\vert$XXXXXX$>$ +1$\vert$XXXYXX$>$ +1$\vert$XXYXXX$>$ +1$\vert$XYXXXX$>$ +1$\vert$XYXYXX$>$ +1$\vert$YXXXXX$>$ +1$\vert$YXXYXX$>$ +1$\vert$YXYXXX$>$ & &  \\ 
\cline{2-3}
\hline
Dyck 2 & 2 & 1$\vert$YX$>$ & 2& unknown \\ 
\cline{2-3}
 & 4 & 1$\vert$YXYX$>$ +1$\vert$YYXX$>$ & &  \\ 
\cline{2-3}
 & 6 & 1$\vert$YXYXYX$>$ +1$\vert$YXYYXX$>$ +1$\vert$YYXXYX$>$ +1$\vert$YYXYXX$>$ +1$\vert$YYYXXX$>$ & &  \\ 
\cline{2-3}
\hline
Dyck 1 & 2 & 1$\vert$XX$>$ & 2& unknown \\ 
\cline{2-3}
 & 4 & 1$\vert$XYXX$>$ & &  \\ 
\cline{2-3}
 & 6 & 1$\vert$XXYYXX$>$ +1$\vert$XYXYXX$>$ & &  \\ 
\cline{2-3}
\hline
Majumdar-Ghosh & 2 & 1$\vert$XY$>$ +-1$\vert$YX$>$ & $\infty$& \textbf{unknown} \\ 
\cline{2-3}
 & 4 & 1$\vert$XYXY$>$ +-1$\vert$XYYX$>$ +-1$\vert$YXXY$>$ +1$\vert$YXYX$>$ & &  \\ 
\cline{2-3}
 & 6 & 1$\vert$XYXYXY$>$ +-1$\vert$XYXYYX$>$ +-1$\vert$XYYXXY$>$ +1$\vert$XYYXYX$>$ +-1$\vert$YXXYXY$>$ +1$\vert$YXXYYX$>$ +1$\vert$YXYXXY$>$ +-1$\vert$YXYXYX$>$ & &  \\ 
\cline{2-3}
\hline
AKLT & 2 & 1$\vert$XY$>$ +-1$\vert$YX$>$ & $\infty$& \textbf{unknown} \\ 
\cline{2-3}
 & 4 & 1$\vert$XYXY$>$ +-1$\vert$XYYX$>$ +-1$\vert$YXXY$>$ +1$\vert$YXYX$>$ & &  \\ 
\cline{2-3}
 & 6 & 1$\vert$XYXYXY$>$ +-1$\vert$XYXYYX$>$ +-1$\vert$XYYXXY$>$ +1$\vert$XYYXYX$>$ +-1$\vert$YXXYXY$>$ +1$\vert$YXXYYX$>$ +1$\vert$YXYXXY$>$ +-1$\vert$YXYXYX$>$ & &  \\ 
\cline{2-3}
\hline
\end{tabular}
\caption{\textbf{Target states for quantum circuit design (Part 2)}}
\end{table}

\begin{table}[H]
\begin{tabular}{V{4cm}|V{1cm}|V{12cm}}
State Name & size & Quantum State string\\
\hline
\hline
Linear & 2 & 1$\vert$XX$>$ +1$\vert$XY$>$ +1$\vert$YX$>$ +-1$\vert$YY$>$  \\ 
\cline{2-3}
 & 4 & 1$\vert$XXXX$>$ +1$\vert$XXXY$>$ +1$\vert$XXYX$>$ +-1$\vert$XXYY$>$ +1$\vert$XYXX$>$ +1$\vert$XYXY$>$ +-1$\vert$XYYX$>$ +1$\vert$XYYY$>$ +1$\vert$YXXX$>$ +1$\vert$YXXY$>$ +1$\vert$YXYX$>$ +-1$\vert$YXYY$>$ +-1$\vert$YYXX$>$ +-1$\vert$YYXY$>$ +1$\vert$YYYX$>$ +-1$\vert$YYYY$>$  \\ 
\cline{2-3}
 & 6 & 1$\vert$XXXXXX$>$ +1$\vert$XXXXXY$>$ +1$\vert$XXXXYX$>$ +-1$\vert$XXXXYY$>$ +1$\vert$XXXYXX$>$ +1$\vert$XXXYXY$>$ +-1$\vert$XXXYYX$>$ +1$\vert$XXXYYY$>$ +1$\vert$XXYXXX$>$ +1$\vert$XXYXXY$>$ +1$\vert$XXYXYX$>$ +-1$\vert$XXYXYY$>$ +-1$\vert$XXYYXX$>$ +-1$\vert$XXYYXY$>$ +1$\vert$XXYYYX$>$ +-1$\vert$XXYYYY$>$ +1$\vert$XYXXXX$>$ +1$\vert$XYXXXY$>$ +1$\vert$XYXXYX$>$ +-1$\vert$XYXXYY$>$ +1$\vert$XYXYXX$>$ +1$\vert$XYXYXY$>$ +-1$\vert$XYXYYX$>$ +1$\vert$XYXYYY$>$ +-1$\vert$XYYXXX$>$ +-1$\vert$XYYXXY$>$ +-1$\vert$XYYXYX$>$ +1$\vert$XYYXYY$>$ +1$\vert$XYYYXX$>$ +1$\vert$XYYYXY$>$ +-1$\vert$XYYYYX$>$ +1$\vert$XYYYYY$>$ +1$\vert$YXXXXX$>$ +1$\vert$YXXXXY$>$ +1$\vert$YXXXYX$>$ +-1$\vert$YXXXYY$>$ +1$\vert$YXXYXX$>$ +1$\vert$YXXYXY$>$ +-1$\vert$YXXYYX$>$ +1$\vert$YXXYYY$>$ +1$\vert$YXYXXX$>$ +1$\vert$YXYXXY$>$ +1$\vert$YXYXYX$>$ +-1$\vert$YXYXYY$>$ +-1$\vert$YXYYXX$>$ +-1$\vert$YXYYXY$>$ +1$\vert$YXYYYX$>$ +-1$\vert$YXYYYY$>$ +-1$\vert$YYXXXX$>$ +-1$\vert$YYXXXY$>$ +-1$\vert$YYXXYX$>$ +1$\vert$YYXXYY$>$ +-1$\vert$YYXYXX$>$ +-1$\vert$YYXYXY$>$ +1$\vert$YYXYYX$>$ +-1$\vert$YYXYYY$>$ +1$\vert$YYYXXX$>$ +1$\vert$YYYXXY$>$ +1$\vert$YYYXYX$>$ +-1$\vert$YYYXYY$>$ +-1$\vert$YYYYXX$>$ +-1$\vert$YYYYXY$>$ +1$\vert$YYYYYX$>$ +-1$\vert$YYYYYY$>$  \\ 
\cline{2-3}
\hline
Ring & 2 & 1$\vert$XX$>$ +1$\vert$XY$>$ +1$\vert$YX$>$ +1$\vert$YY$>$  \\ 
\cline{2-3}
 & 4 & 1$\vert$XXXX$>$ +1$\vert$XXXY$>$ +1$\vert$XXYX$>$ +-1$\vert$XXYY$>$ +1$\vert$XYXX$>$ +1$\vert$XYXY$>$ +-1$\vert$XYYX$>$ +1$\vert$XYYY$>$ +1$\vert$YXXX$>$ +-1$\vert$YXXY$>$ +1$\vert$YXYX$>$ +1$\vert$YXYY$>$ +-1$\vert$YYXX$>$ +1$\vert$YYXY$>$ +1$\vert$YYYX$>$ +1$\vert$YYYY$>$  \\ 
\cline{2-3}
 & 6 & 1$\vert$XXXXXX$>$ +1$\vert$XXXXXY$>$ +1$\vert$XXXXYX$>$ +-1$\vert$XXXXYY$>$ +1$\vert$XXXYXX$>$ +1$\vert$XXXYXY$>$ +-1$\vert$XXXYYX$>$ +1$\vert$XXXYYY$>$ +1$\vert$XXYXXX$>$ +1$\vert$XXYXXY$>$ +1$\vert$XXYXYX$>$ +-1$\vert$XXYXYY$>$ +-1$\vert$XXYYXX$>$ +-1$\vert$XXYYXY$>$ +1$\vert$XXYYYX$>$ +-1$\vert$XXYYYY$>$ +1$\vert$XYXXXX$>$ +1$\vert$XYXXXY$>$ +1$\vert$XYXXYX$>$ +-1$\vert$XYXXYY$>$ +1$\vert$XYXYXX$>$ +1$\vert$XYXYXY$>$ +-1$\vert$XYXYYX$>$ +1$\vert$XYXYYY$>$ +-1$\vert$XYYXXX$>$ +-1$\vert$XYYXXY$>$ +-1$\vert$XYYXYX$>$ +1$\vert$XYYXYY$>$ +1$\vert$XYYYXX$>$ +1$\vert$XYYYXY$>$ +-1$\vert$XYYYYX$>$ +1$\vert$XYYYYY$>$ +1$\vert$YXXXXX$>$ +-1$\vert$YXXXXY$>$ +1$\vert$YXXXYX$>$ +1$\vert$YXXXYY$>$ +1$\vert$YXXYXX$>$ +-1$\vert$YXXYXY$>$ +-1$\vert$YXXYYX$>$ +-1$\vert$YXXYYY$>$ +1$\vert$YXYXXX$>$ +-1$\vert$YXYXXY$>$ +1$\vert$YXYXYX$>$ +1$\vert$YXYXYY$>$ +-1$\vert$YXYYXX$>$ +1$\vert$YXYYXY$>$ +1$\vert$YXYYYX$>$ +1$\vert$YXYYYY$>$ +-1$\vert$YYXXXX$>$ +1$\vert$YYXXXY$>$ +-1$\vert$YYXXYX$>$ +-1$\vert$YYXXYY$>$ +-1$\vert$YYXYXX$>$ +1$\vert$YYXYXY$>$ +1$\vert$YYXYYX$>$ +1$\vert$YYXYYY$>$ +1$\vert$YYYXXX$>$ +-1$\vert$YYYXXY$>$ +1$\vert$YYYXYX$>$ +1$\vert$YYYXYY$>$ +-1$\vert$YYYYXX$>$ +1$\vert$YYYYXY$>$ +1$\vert$YYYYYX$>$ +1$\vert$YYYYYY$>$  \\ 
\cline{2-3}
\hline
Star & 2 & 1$\vert$XX$>$ +1$\vert$XY$>$ +1$\vert$YX$>$ +-1$\vert$YY$>$  \\ 
\cline{2-3}
 & 4 & 1$\vert$XXXX$>$ +1$\vert$XXXY$>$ +1$\vert$XXYX$>$ +-1$\vert$XXYY$>$ +1$\vert$XYXX$>$ +-1$\vert$XYXY$>$ +1$\vert$XYYX$>$ +1$\vert$XYYY$>$ +1$\vert$YXXX$>$ +-1$\vert$YXXY$>$ +1$\vert$YXYX$>$ +1$\vert$YXYY$>$ +1$\vert$YYXX$>$ +1$\vert$YYXY$>$ +1$\vert$YYYX$>$ +-1$\vert$YYYY$>$  \\ 
\cline{2-3}
 & 6 & 1$\vert$XXXXXX$>$ +1$\vert$XXXXXY$>$ +1$\vert$XXXXYX$>$ +-1$\vert$XXXXYY$>$ +1$\vert$XXXYXX$>$ +-1$\vert$XXXYXY$>$ +1$\vert$XXXYYX$>$ +1$\vert$XXXYYY$>$ +1$\vert$XXYXXX$>$ +-1$\vert$XXYXXY$>$ +1$\vert$XXYXYX$>$ +1$\vert$XXYXYY$>$ +1$\vert$XXYYXX$>$ +1$\vert$XXYYXY$>$ +1$\vert$XXYYYX$>$ +-1$\vert$XXYYYY$>$ +1$\vert$XYXXXX$>$ +-1$\vert$XYXXXY$>$ +1$\vert$XYXXYX$>$ +1$\vert$XYXXYY$>$ +1$\vert$XYXYXX$>$ +1$\vert$XYXYXY$>$ +1$\vert$XYXYYX$>$ +-1$\vert$XYXYYY$>$ +1$\vert$XYYXXX$>$ +1$\vert$XYYXXY$>$ +1$\vert$XYYXYX$>$ +-1$\vert$XYYXYY$>$ +1$\vert$XYYYXX$>$ +-1$\vert$XYYYXY$>$ +1$\vert$XYYYYX$>$ +1$\vert$XYYYYY$>$ +1$\vert$YXXXXX$>$ +-1$\vert$YXXXXY$>$ +1$\vert$YXXXYX$>$ +1$\vert$YXXXYY$>$ +1$\vert$YXXYXX$>$ +1$\vert$YXXYXY$>$ +1$\vert$YXXYYX$>$ +-1$\vert$YXXYYY$>$ +1$\vert$YXYXXX$>$ +1$\vert$YXYXXY$>$ +1$\vert$YXYXYX$>$ +-1$\vert$YXYXYY$>$ +1$\vert$YXYYXX$>$ +-1$\vert$YXYYXY$>$ +1$\vert$YXYYYX$>$ +1$\vert$YXYYYY$>$ +1$\vert$YYXXXX$>$ +1$\vert$YYXXXY$>$ +1$\vert$YYXXYX$>$ +-1$\vert$YYXXYY$>$ +1$\vert$YYXYXX$>$ +-1$\vert$YYXYXY$>$ +1$\vert$YYXYYX$>$ +1$\vert$YYXYYY$>$ +1$\vert$YYYXXX$>$ +-1$\vert$YYYXXY$>$ +1$\vert$YYYXYX$>$ +1$\vert$YYYXYY$>$ +1$\vert$YYYYXX$>$ +1$\vert$YYYYXY$>$ +1$\vert$YYYYYX$>$ +-1$\vert$YYYYYY$>$  \\ 
\cline{2-3}
\hline
\end{tabular}
\caption{\textbf{Target states for Quantum Graph States}}
\end{table}


\end{document}